\documentclass[12pt,a4paper]{article}
\pdfoutput=1
\usepackage{cite}
\usepackage{jheppub}
\usepackage{amsmath,amsthm,amssymb,graphicx,slashed}
\usepackage{booktabs}
\usepackage{hyperref}
\usepackage[usenames,dvipsnames]{xcolor}

\numberwithin{equation}{section}

\usepackage{color}
\usepackage{amsmath}
\usepackage{amsfonts}
\usepackage{verbatim}
\usepackage{amssymb}
\usepackage{mathrsfs}
\usepackage{cite}
\usepackage{graphicx}
\providecommand{\U}[1]{\protect\rule{.1in}{.1in}}

\title{{\huge {Trace anomaly and Counterterms in Designer Gravity}}}

\author[1]{Andr\'{e}s Anabal\'{o}n,}
\author[2,3]{Dumitru Astefanesei,}
\author[3,4]{David Choque,}
\author[5]{ and Cristi\'an Mart\'{\i}nez}

\affiliation[1]{Departamento de Ciencias, Facultad de Artes Liberales and
Facultad de Ingenier\'{\i}a y Ciencias, Universidad
Adolfo Ib\'{a}\~{n}ez, Av. Padre Hurtado 750, Vi\~{n}a del Mar,
Chile}
\affiliation[2]{Instituto de F\'\i sica, Pontificia
Universidad Cat\'olica de Valpara\'\i so, Casilla 4059,
Valpara\'{\i}so, Chile}

\affiliation[3]{Max-Planck-Institut
f\"ur Gravitationsphysik,
Albert-Einstein-Institut, 14476 Golm, Germany}

\affiliation[4]{Universidad T\'{e}cnica Federico Santa Mar\'{\i}a,
 Av. Espa\~{n}a 1680, Valpara\'{\i}so, Chile}

\affiliation[5]{Centro de Estudios Cient\'{\i}ficos (CECs),
 Av. Arturo Prat 514, Valdivia, Chile}

\emailAdd{andres.anabalon@uai.cl}
\emailAdd{dumitru.astefanesei@ucv.cl}
\emailAdd{brst1010123@gmail.com}
\emailAdd{martinez@cecs.cl}

\abstract{
We construct concrete counterterms of the Balasubramanian-Kraus type for
Einstein-scalar theories with designer gravity boundary conditions in
AdS$_{4}$, so that the total action is finite on-shell and satisfy a well
defined variational principle. We focus on scalar fields with the conformal
mass $m^{2}=-2l^{-2}$ and show that the holographic mass matches the
Hamiltonian mass for any boundary conditions. We compute the trace anomaly 
of the dual field theory in the generic case, as well as when there exist
logarithmic branches of non-linear origin. As expected, the anomaly
vanishes for the boundary conditions that are AdS invariant. When the anomaly
does not vanish, the dual stress tensor describes a thermal gas with an
equation of state related to the boundary conditions of the scalar field. In
the case of a vanishing anomaly, we recover the dual theory of a massless
thermal gas. As an application of the formalism, we consider a general family
of exact hairy black hole solutions that, for some particular values of the
parameters in the moduli potential, contains solutions of four-dimensional
gauged $\mathcal{N}=8$ supergravity and its $\omega$-deformation. Using the
AdS/CFT duality dictionary, they correspond to triple trace deformations of
the dual field theory.
}

\begin{document}

\maketitle

\section{Introduction}

A well-known check of the AdS/CFT duality \cite{Maldacena:1997re} is the exact
matching of the AdS$_{5}$ vacuum energy and the Casimir energy of the large
$N$ limit of $SU(N)$, $\mathcal{N}=4$ super Yang-Mills on $S^{3}$
\cite{Henningson:1998gx,Balasubramanian:1999re}. Technically speaking, this
result was implemented by fixing the boundary data, namely imposing Dirichlet
boundary conditions. Indeed, AdS is not globally hyperbolic, which means that
besides initial conditions it is necessary to also provide boundary conditions
for the evolution of a given field so that is well-defined. This was analyzed
in-extenso for the linearized dynamics of spin-$0$, $1$, and $2$ fields in AdS
\cite{Ishibashi:2004wx}, where, in particular, all the self-adjoint extensions
for the relevant spin-$0$ Sturm-Liouville operators were found for scalar
fields of mass $m$, which satisfy
\begin{equation}
\label{BFi}m_{BF}^{2}+\frac{1}{l^{2}}>m^{2}\geq m_{BF}^{2} \,\,\,\, , \qquad
m_{BF}^{2}=-\frac{(D-1)^{2}}{4l^{2}}%
\end{equation}
where $l$ is the AdS radius, $D$ is the spacetime dimension, and $m_{BF}^{2}$
is the Breitenlohner-Freedman (BF) bound \cite{BF}. Furthermore, the
backreaction of these generalized boundary conditions was considered, and its
contribution to the spacetime energy has been computed in different ways
\cite{Henneaux:2002wm,Barnich:2002pi, Henneaux:2004zi, Hertog:2004dr,
Hertog:2004ns, Henneaux:2006hk, Amsel:2006uf,Anabalon:2014fla}.

One of the interesting outputs of \cite{Henneaux:2006hk} is the existence of
logarithmic branches of non-linear origin at certain values of the scalar
field mass. In particular, this occurs when the scalar field mass is the one
of the scalars of four-dimensional gauged $\mathcal{N}=8$ supergravity, namely
$m^{2}=-2l^{-2}$. It is interesting to note that the interpretation of the AdS
invariant boundary conditions as a multi-trace deformation of the dual theory
was given before this exhaustive analysis in \cite{Witten:2001ua}(see, also,
the nice work \cite{Aharony:2015afa}). Some of these theories are equivalently
defined by the requirement of the existence of a soliton with a given value of
the scalar field at the origin. Such constructions go by the name of designer
gravity \cite{Hertog:2004ns}. Techniques for constructing \textit{exact} hairy
black hole solutions that are relevant for our work can be found
in\cite{Acena:2012mr, Acena:2013jya, Anabalon:2013sra, Fan:2015ykb}.

In \cite{Balasubramanian:1999re,Skenderis:2000in,Skenderis:2002wp},
the energy of the gravitational configuration is connected to the
gauge theory through the quasi-local Brown-York energy momentum
tensor \cite{Brown:1992br}. A non-trivial contribution to the mass
when the boundary conditions break the conformal symmetry is
expected to imply a modification of the trace anomaly, as was shown
to be the case using the Hamilton-Jacobi equation in
\cite{Papadimitriou:2007sj}. However, to the best of our knowledge,
the same construction has not been done in the spirit of the work by 
Balasubramanian and Kraus \cite{Balasubramanian:1999re}, neither 
extended to the logarithmic
branches of non-linear origin at $m^{2}=-2l^{-2}$, which is the main
objective of this paper.\footnote{In three dimensions, similar work was 
done in \cite{Aparicio:2012yq} 
and for the particular case of 
Dirichlet boundary conditions see, e.g., \cite{Nojiri:1998dh, Nojiri:2000kh}.} 
We use the Hamiltonian formalism as a guide for constructing the counterterms. 

It is important 
to emphasize that, when the
conformal symmetry is broken in the boundary, the mass of Ashtekar-Magnon-Das
(AMD)\cite{Ashtekar:1999jx} does not match the Hamiltonian mass
\cite{Anabalon:2014fla} and so it also does not match the holographic mass.
Therefore, for AdS black hole solutions when the conformal symmetry is broken
in the boundary, e.g \cite{Chow:2013gba}, the right mass is not the AMD mass. As 
a concrete application, we shall consider a relatively simple family of
hairy black hole solutions \cite{Anabalon:2013eaa} that, however, is general
enough to include the single scalar field truncations of four-dimensional
gauged $\mathcal{N}=8$ supergravity as well as its $\omega$-deformation
\cite{Dall'Agata:2012bb} (see, also, \cite{Tarrio:2013qga, Dibitetto:2014sfa,
Gallerati:2014xra}).

The remainder of the paper is organized as follows: in Section 2 we review
mixed AdS boundary conditions in the context of AdS/CFT duality. In Section 3
we provide the counterterms that regularize the action and verify that the
variational principle is well defined when the action is supplemented with
these counterterms. We compute the free energy of a generic hairy black hole
solution with mixed boundary conditions of the scalar field and the relevant
thermodynamical quantities. Section 4 contains the general formalism for
computing the regularized quasilocal stress tensor (for both, the logarithmic
and non-logarithmic branches). Using the AdS/CFT duality dictionary, we then
compute the stress tensor of the dual field theory and the anomaly when the
boundary conditions break conformal symmetry. In Section 5 we compare
different types of constructions of gravitational mass in AdS for mixed
boundary conditions of the scalar field. When the conformal symmetry is broken
the holographic and Hamiltonian mass match, but do not match the AMD mass.
Then, we work out concrete examples of hairy black hole solutions that are
dual to triple trace deformations of the boundary field theory. Finally, we
end with some conclusions and future directions.

\section{General AdS$_{4}$ boundary conditions and multi-trace deformations in
the dual theory}

\label{sec2}

In this section, we review the role of AdS boundary conditions in the context
of AdS/CFT duality \cite{Maldacena:1997re}. According to `holographic'
dictionary, imposing mixed boundary conditions on the scalar field (in the
bulk) corresponds to perturbing the large $N$ boundary theory by a relevant,
irrelevant or marginal multi-trace deformation \cite{Witten:2001ua}.

Let us start by exhibiting some known facts about the AdS/CFT duality
\cite{Maldacena:1997re}. We would like to describe what kind of boundary
conditions preserve the conformal symmetry of the dual field theory and
interpret them in the context of the AdS/CFT duality \cite{Witten:2001ua,
Hertog:2004dr, Henneaux:2006hk}. First, we describe the AdS$_{4}$ spacetime
and explain how the symmetries of the two dual theories match. That is, the
isometry group $SO(3,2)$ of AdS$_{4}$ acts on the (conformal) boundary as the
conformal group\footnote{The conformal group of Minkowski spacetime is the
invariance group of the light cone, in other words all the transformations
that leave $ds^{2}=0$ invariant.} acting on Minkowski spacetime.

AdS spacetime has the maximal number of isometries in every dimension. Hence,
it has a simple form in a large number of coordinate systems (see, e.g.,
\cite{Emparan:1999pm} for a discussion in the context of AdS/CFT duality).
Depending on the choice of the radial coordinate, the slices at constant
radius can have a different geometry or even a different topology. For
example, one can foliate AdS$_{4}$ with the following slices:
\begin{equation}
d\bar{s}^{2}=\bar{g}_{\mu\nu}dx^{\mu}dx^{\nu}=-\left(  k+\frac{r^{2}}{l^{2}}
\right)  dt^{2} + \frac{dr^{2}}{k+\frac{r^{2}}{l^{2}}}+r^{2} d\Sigma^{2}_{k}
\label{metricAdS}%
\end{equation}
where $k=\{+1, 0, -1\}$ for the spherical ($d\Sigma_{1}^{2}=d\Omega^{2}$),
toroidal ($d\Sigma_{0}^{2} = dx^{2}+dy^{2}$), and hyperbolic ($d\Sigma_{-1} =
dH^{2}$) foliations, respectively. Here, $d\Omega^{2}$ and $dH^{2}$ are the
unit metrics on the $2$-dimensional sphere and hyperboloid, respectively. The
radius $l$ of AdS$_{4}$ is related to the cosmological constant by $\Lambda=
-3/l^{2}$.

The conformal boundary is at $r \rightarrow\infty$, for which the induced
metric is
\begin{equation}
h_{ab}dx^{a}dx^{b}=\frac{r^{2}}{l^{2}}(-dt^{2}+l^{2}d\Sigma_{k}^{2})
\end{equation}
and now it is clear that the background geometry where the field theory lives
is related to the boundary geometry by a conformal transformation. Therefore,
a bulk metric is associated with a conformal structure at infinity. The
conformal factor is going to play an important role when we are going to
compute the boundary stress tensor.

Even if different foliations of AdS$_{4}$ are related by local coordinate
transformations, the corresponding dual gauge theories are physically
inequivalent (for example, in the $k=1$ case there is a Hawking-Page phase
transition, but not for $k=0$). This is due to the fact that different
spacelike foliations of the background geometry lead to different definitions
of the time coordinate (and so the Hamiltonian) of the dual quantum system.

Starting with $k=0$ form of AdS$_{4}$ metric and using the change of
coordinates $r=l^{2}/z$, we obtain
\begin{equation}
\label{AdSz}d\bar{s}^{2}=\frac{l^{2}}{z^{2}}(dz^{2} - dt^{2} + dx^{2} +
dy^{2})
\end{equation}
In these coordinates, which cover only part of AdS$_{4}$ spacetime, the
Minkowski spacetime appears naturally as the conformal boundary. The finite
isometries of AdS$_{4}$ map the boundary $z=0$ to itself and, moreover, act as
conformal transformations in the boundary. In particular, the transformation
$(z, t, x, y)\leftarrow\lambda(z,t,x,y)$, which leaves the metric (\ref{AdSz})
invariant, acts as the dilation (scale transformation) in the boundary. Since
the AdS spacetime is not globally hyperbolic, one has to impose boundary
conditions. Within the AdS/CFT duality, various deformations of the AdS
boundary conditions are interpreted as dual to deformations of the CFT. It is
well known \cite{BF, Ishibashi:2004wx} that a scalar of arbitrary mass in AdS
can have both normalizable and non-normalizable modes. It was shown in
\cite{Balasubramanian:1998sn, Balasubramanian:1998de} that the normalizable
modes describe fluctuations in the bulk and the non-normalizable modes
correspond to operator insertions in the boundary dual field theory. We are
interested in the case when both modes are normalizable:
\begin{equation}
m^{2}_{BF} + \frac{1}{l^{2}}> m^{2} \geq m^{2}_{BF} \, ,
\,\,\,\,\,\,\,\,\,\,\,\,\,\,\,\,\, m^{2}_{BF}= -\frac{9}{4l^{2}}%
\end{equation}
where $m^{2}_{BF}$ is the BF bound (\ref{BFi}) in four dimensions.

In what follows we briefly review the boundary conditions that accommodate a
scalar field whose mass corresponds to the conformal one. We are interested in
the action
\begin{equation}
I[g_{\mu\nu},\phi] = \int_{\mathcal{M}}{d^{4}x\sqrt{-g}\biggl{[}\frac
{R}{2\kappa}-\frac{1}{2}(\partial\phi)^{2}-V(\phi)\biggr{]}} + \frac{1}%
{\kappa}\int_{\partial\mathcal{M}}{d^{3}xK\sqrt{-h}} \label{action}%
\end{equation}
where $V(\phi)$ is the scalar potential, $\kappa=8\pi G $ with $G$ the
Newton gravitational constant, and the last term is the Gibbons-Hawking
boundary term. Here, $h$ is the determinant of the boundary metric and $K$ is
the trace of the extrinsic curvature. The equations of motion for the scalar
field and metric are
\begin{equation}
\frac{1}{\sqrt{-g}}\partial_{\mu}\left(  \sqrt{-g}g^{\mu\nu}\partial_{\nu}%
\phi\right)  -\frac{\partial V}{\partial\phi}=0 \label{dil}%
\end{equation}%
\begin{equation}
E_{\mu\nu}=R_{\mu\nu}-\frac{1}{2}g_{\mu\nu}R-\kappa T_{\mu\nu}^{\phi}=0
\label{eqmotion}%
\end{equation}
where the stress tensor of the scalar field is
\begin{equation}
T_{\mu\nu}^{\phi}=\partial_{\mu}\phi\partial_{\nu}\phi-g_{\mu\nu}\left[
\frac{1}{2}\left(  \partial\phi\right)  ^{2}+V(\phi)\right]
\end{equation}
We work with the general ansatz
\begin{equation}
ds^{2}=-N(r)dt^{2}+H(r)dr^{2}+S(r)d\Sigma_{k}^{2} \label{Ansatz1}%
\end{equation}
As it was shown first in three dimensions \cite{Henneaux:2002wm}, and then
generalized in four and higher dimensions \cite{Henneaux:2004zi,Hertog:2004dr,
Henneaux:2006hk,Amsel:2006uf}, in the presence of the scalar fields the
standard AdS boundary conditions are modified. One can obtain the right
fall-off for the $g_{rr}$ component of the metric by considering the equations
of motion and using the fall-off of the scalar field. A general discussion for
any mass of the scalar field in the range (\ref{BFi}) can be found in
\cite{Henneaux:2006hk}, but in this work we focus on the concrete case of the
conformal mass in four dimensions $m^{2}=-2l^{-2}$. We start with the
potential
\begin{equation}
V(\phi)=-\frac{3}{\kappa l^{2}}-\frac{\phi^{2}}{l^{2}} +O(\phi^{4})
\label{vphinolog}%
\end{equation}
The fall-off of the scalar field in this case is
\begin{equation}
\phi(r)=\frac{\alpha}{r}+\frac{\beta}{r^{2}}+O(r^{-3}) \label{phi}%
\end{equation}
In order to accommodate the black hole of Sec. \ref{exact3} we consider the
following asymptotic behavior for the $N(r)$ and $S(r)$ metric coefficients
\begin{align}
N(r)  &  =-g_{tt}=\frac{r^{2}}{l^{2}}+k-\frac{\mu}{r}+O(r^{-2})\label{gtt}\\
S(r)  &  =r^{2}+O(r^{-2}) \label{gSS}%
\end{align}
Now, we use the combination of the equations of motion (\ref{eqmotion}),
$E_{t}^{t}-E_{r}^{r}=0$, from which we obtain
\begin{equation}
NS^{^{\prime}2}H-2NS^{^{\prime\prime}}HS+(NH)^{^{\prime}}S^{^{\prime}%
}S-2\kappa NHS^{2}\phi^{^{\prime}2}=0
\end{equation}
and then
\begin{equation}
H(r)=g_{rr}=\frac{l^{2}}{r^{2}}+\frac{l^{4}}{r^{4}}\biggl{(}-k-\frac
{\alpha^{2}\kappa}{2l^{2}}\biggr{)}+\frac{l^{5}}{r^{5}}\biggl{(}\frac{\mu}%
{l}-\frac{4\kappa\alpha\beta}{3l^{3}}\biggr{)}+O(r^{-6})
\end{equation}
The reason we would like to obtain the fall-off of $g_{rr}$ in this way is
because the Hamiltonian mass can be read off from it --- if there is a
contribution of the scalar field to the mass, one should be able to identify
it in $g_{rr}$.

From now on, we use the generic notation for the expansion of $g_{rr}$ as
\begin{equation}
g_{rr}=\frac{l^{2}}{r^{2}}+\frac{al^{4}}{r^{4}}+\frac{bl^{5}}{r^{5}}+O(r^{-6})
\label{grr}%
\end{equation}
where $a=-k-\frac{\kappa\alpha^{2}}{2l^{2}}$ and $b=\frac{\mu}{l}%
-\frac{4\kappa\alpha\beta}{3l^{3}}$. At this point, it is interesting to
investigate when the asymptotic conditions are AdS invariant and the
Hamiltonian is well defined. It seems that, for some special functional
relationship on the coefficients $\alpha$ and $\beta$ of the modes of the
scalar field, both conditions are satisfied. This was explicitly done in
\cite{Hertog:2004dr, Henneaux:2006hk} and here we just present the result:
\begin{equation}
\beta=C\alpha^{2} \label{betanon}%
\end{equation}
Interestingly enough, one can also obtain a finite Hamiltonian when the
boundary conformal symmetry is broken.

A similar analysis can be done for the so-called \textit{logarithmic branch}
\cite{Henneaux:2004zi}. In what follows we would like to carefully analyze
this case and present details we are going to use in the next sections.

It is well known that a second order differential equation has two linearly
independent solutions. When the ratio of the roots of the indicial equation is
an integer, the solution may develop a logarithmic branch. This is exactly
what happens when the scalar field saturates the BF bound, in which case the
leading fall-off contains a logarithmic term \cite{Henneaux:2004zi}. However,
we are interested in a scalar field with the conformal mass $m^{2}=-2l^{-2}$.
To obtain the logarithmic branch, a cubic term in the asymptotic expansion of
the scalar field potential is necessary \cite{Henneaux:2006hk}
\begin{equation}
V(\phi)=-\frac{3}{\kappa l^{2}}-\frac{\phi^{2}}{l^{2}}+\lambda\phi^{3}%
+O(\phi^{4}) \label{vphilog}%
\end{equation}
so that the fall-off of the scalar field to be considered is
\begin{equation}
\phi(r)=\frac{\alpha}{r}+\frac{\beta}{r^{2}}+\frac{\gamma\ln(r)}{r^{2}%
}+O(r^{-3}) \label{philog}%
\end{equation}
To obtain the fall-off of $g_{rr}$ we use the same fall-off for the other
components of the metric and the same combination of the equations of motion
as in the non-logarithmic branch, $E_{t}^{ t}-E_{r}^{ r}=0$. We get
\begin{equation}
H(r)=g_{rr}=\frac{l^{2}}{r^{2}}+\frac{l^{4}}{r^{4}}\biggl{(}-k-\frac
{\kappa\alpha^{2}}{2l^{2}}\biggr{)}+\frac{l^{5}}{r^{5}}\biggl{(}\frac{\mu}%
{l}-\frac{4\kappa\alpha\beta}{3l^{3}}+\frac{2\kappa\alpha\gamma}{9l^{3}%
}\biggr{)}+\frac{l^{5}\ln{r}}{r^{5}}\biggl{(}-\frac{4\kappa\alpha\gamma
}{3l^{3}}\biggr{)}+O\biggl{[}\frac{\ln{(r)^{2}}}{r^{6}}\biggr{]}
\label{grrlog}%
\end{equation}
Using again the generic notation for the asymptotic expansion of $g_{rr}$
\begin{equation}
H(r)=\frac{l^{2}}{r^{2}}+\frac{l^{4}a}{r^{4}}+\frac{l^{5}b}{r^{5}}+\frac
{l^{5}c\ln{r}}{r^{5}}+O\biggl{[}\frac{\ln{(r)^{2}}}{r^{6}}\biggr{]}
\end{equation}
we identify the relevant coefficients as
\begin{align}
a  &  =-k-\frac{\alpha^{2}\kappa}{2l^{2}}; \qquad b=\frac{\mu}{l}%
-\frac{4\kappa\alpha\beta}{3l^{3}}+\frac{2\kappa\alpha\gamma}{9l^{3}}; \qquad
c=-\frac{4\kappa\gamma\alpha}{3l^{3}} \label{abglog}%
\end{align}
Now, let us check when the fall-off of the scalar field we have considered is
compatible with its equation of motion:
\begin{equation}
\partial_{r}\biggl{(}\frac{\phi^{^{\prime}}S\sqrt{N}}{\sqrt{H}}%
\biggr{)}-S\sqrt{NH}\frac{\partial V}{\partial\phi}=0
\end{equation}
In the asymptotic region, $r\rightarrow\infty$, this equation becomes
\begin{equation}
\frac{3\alpha^{2}l^{2}\lambda+\gamma}{l^{2}}+O(r^{-1})=0
\end{equation}
and so the coefficient $\gamma$ is fixed by $\alpha$ as $\gamma=-3l^{2}%
\lambda\alpha^{2}$ (or, using the notation that we shall use below,
$\gamma=C_{\gamma}\alpha^{2}$, where $C_{\gamma}=-3l^{2}\lambda$). This result
is important because, as we will see shortly, is also part of the conditions
that preserve the conformal symmetry of the boundary.

The last step in our derivation is to investigate when the boundary conditions
are preserved under the asymptotic AdS symmetry. The corresponding asymptotic
Killing vector $\xi^{\mu}=(\xi^{r},\xi^{m})$ is
\begin{align}
\xi^{r}  &  =r\eta^{r}(x^{m})+O(r^{-1})\\
\xi^{m}  &  =O(1)\nonumber
\end{align}
where $\{m\}$ is an index that run over the time an angular coordinates. The
fall-off of the scalar field should be invariant under the asymptotic AdS
symmetries and so we obtain:
\begin{equation}
\label{seriesphi}\phi^{\prime}(x)=\phi(x)+\xi^{\mu}\partial_{\mu}\phi(x)=
\frac{\bar{\alpha}}{r}+\frac{\bar{\beta}}{r^{2}}+\frac
{\bar{\gamma} \ln(r)}{r^{2}}+O(r^{-3})
\end{equation}
where
\begin{align}
\bar{\alpha}=  &  \alpha-\eta^{r}\alpha+\xi^{m}\partial_{m}\alpha\\
\bar{\beta}=  &  \beta-\eta^{r}(2\beta-\gamma)+\xi^{m}\partial_{m}%
\beta\nonumber\\
\bar{\gamma}=  &  \gamma-2\gamma\eta^{r}+\xi^{m}\partial_{m}%
\gamma\nonumber\\
\nonumber
\end{align}
If the coefficients in the series (\ref{seriesphi}) are functionally related, the conformal
symmetry on the boundary fixes the functional relation between the
coefficients so that the equations above are compatible. Hence, one performs a
Taylor expansion of $\bar{\gamma}$ and $\bar{\beta}$ to linear order in $\eta^{r}$ and $\xi^{m}$ to obtain:
\begin{equation}
\eta^{r}%
\biggl{(}2\gamma-\alpha\frac{\partial\gamma}{\partial\alpha}\biggr{)}+\xi
^{m}\biggl{(}\frac{\partial\alpha}{\partial x^{m}}\frac{\partial\gamma
}{\partial\alpha}-\frac{\partial\gamma}{\partial x^{m}}\biggr{)} \label{gamma}=0
\end{equation}
and
\begin{equation}
\eta^{r}\biggl{(}2\beta
-\gamma-\alpha\frac{\partial\beta}{\partial\alpha}\biggr{)}+\xi^{m}%
\biggl{(}\frac{\partial\alpha}{\partial x^{m}}\frac{\partial\beta}%
{\partial\alpha}-\frac{\partial\beta}{\partial x^{m}}\biggr{)} \label{beta}=0 .
\end{equation}

Using the fact that $\eta^{r}$ and $\xi^{m}$ are independent, we get from
(\ref{gamma}) that $2\gamma=\alpha\frac{\partial\gamma}{\partial\alpha}$,
which implies that $\gamma=C_{\gamma}\alpha^{2}$. This is the result obtained
before from the equation of motion for the scalar field. From the integration
of (\ref{beta}) we obtain
\begin{equation}
\beta(\alpha)=(-C_{\gamma}\ln(\alpha)+C)\alpha^{2}%
\end{equation}
When $C_{\gamma}=0$ this result matches the condition found for the
non-logarithmic branch (\ref{betanon}). Again, one can obtain a finite
Hamiltonian even if the conformal invariance is broken.

A precise formulation of the AdS/CFT duality \cite{Maldacena:1997re} was
proposed in \cite{Gubser:1998bc, Witten:1998qj} and developed for multi-trace
deformations in \cite{Witten:2001ua}. The observables in the field theory side
of the duality are the correlation functions of gauge invariant operators,
which are composites of the elementary fields. Any supergravity field $\phi$
corresponds to an operator $O$ in the (boundary) field theory. The duality
relates the generating functional for correlation functions of the operator
$O$ with the string/gravity partition function on AdS space with the boundary
conditions that are imposed on the excitations in the bulk. In our case, the
relevant fields in the bulk are the graviton (metric perturbations) and scalar
field. The corresponding operators in the dual field theory are the
stress-energy tensor $T_{\mu\nu}$ of the dual field theory and a scalar
operator of dimension $\Delta$, respectively.

Let us consider a massive scalar field. By solving the equation of motion
close to the boundary, we obtain:
\begin{equation}
\phi(r)=\frac{\alpha}{r^{\Delta_{-}}}+\frac{\beta}{r^{\Delta_{+}}}+...
\end{equation}
where $\alpha$ and $\beta$ are the leading and sub-leading components of the
asymptotic expansion of the scalar field and $\Delta_{\pm}=\frac{3}{2}\pm
\sqrt{\frac{9}{4}+ m^{2}l^{2}}$.

Depending of the value of the mass, the two modes (on the Lorentzian section)
can be divergent or finite. For example, for a positive squared-mass $m^{2}>0$
the mode $\beta$ is divergent in the interior and finite at the boundary and
the mode $\alpha$ is divergent at the boundary but finite in the interior.
Then, the mode $\beta$ corresponds to source currents in the boundary dual
theory. On the other hand, by turning on the mode $\alpha$, the bulk geometry
is modified while the AdS structure near the boundary may be preserved ---
this is the type of deformation we are interested in this work. Since the bulk
gravity solution is changed, one has to perform a linearized analysis around
the \textit{new} background to calculate the correlation functions. This is
exactly what happens when the `vacuum' around which one expands to obtain the
physical quantities is changed. Then, in the dual theory, there is a similar
situation: the dual field theory is expanded around a vacuum with non-trivial
vacuum expectations values (VEV) for the appropriate operators. Indeed, in the
standard AdS/CFT dictionary \cite{Balasubramanian:1998sn,
Balasubramanian:1998de}, a bulk gravity solution with a non-trivial dilaton
corresponds in the dual field theory to the insertion of a source for an
operator with conformal dimension $\Delta_{-}$, VEV $\alpha$, and current
$\beta=J(x)$.

The spectrum of operators in the dual field theory include all the gauge
invariant quantities, namely product of traces of products of fields (or the
sum of such products). Single-trace operators in the field theory may be
identified with single-particle states in AdS, while multiple-trace operators
correspond to multi-particle states. The significance of the multi-trace
deformations from a point of view of the gravity side was investigated in
\cite{Witten:2001ua, Hertog:2004dr}. The mixed boundary conditions play an
important role because they correspond to a deformation of the field theory
action by
\begin{equation}
I_{CFT}\rightarrow I_{CFT} - \int d^{3}xW[\mathcal{O}{(x)}] \label{triple}%
\end{equation}
where $\beta(x)=\frac{dW}{d\alpha(x)}$, and $W$ is fixed by the boundary
conditions of the string theory side.

In section $5$ we are going to apply this general framework to concrete
analytic hairy black hole solutions.

\section{Counterterms and regularized action}

The usual approach to computing thermodynamic quantities of black holes is to
analytically continue in the time coordinate in order to obtain a Euclidean solution of
the Einstein equations (with negative cosmological constant). In this way, the periodicity of the Euclidean time is related to the
temperature of the black hole and the Euclidean action to the thermodynamical
potential (in our case, the free energy). In this section we explicitly
construct counterterms that cancel the divergences of the action for both
logarithmic and non-logarithmic branches and check that the variational
principle is well possed. We apply the counterterm method to compute the free
energy of hairy black holes with a scalar field with the conformal mass
$m^{2}=-2l^{-2}$.

\subsection{Variational principle}

Our goal is to construct counterterms (boundary terms) that regularize the
action so that the variational principle is well-posed. The boundary terms do
not change the equations of motion and so they can be incorporated in the
action. To make our point, let us first consider the action (\ref{action})
when the scalar field is turned off. In this case, the action has just two
terms: the bulk action and the Gibbons-Hawking surface term necessary to
ensure that the Euler-Lagrange variation is well-defined. The gravitational
action computed in this way (even at tree-level) contains divergences that
arise from integrating over the infinite volume of spacetime. In the AdS/CFT
context, the infrared (IR) divergences of gravity are interpreted as
ultraviolet (UV) divergences of the dual CFT. A well understood way of
computing the bulk action without introducing a background is to add local
counterterms into the action, which remove all divergences, leading to a
finite action corresponding to the partition function of the CFT. For pure AdS
gravity in four dimensions, the action should be supplemented with the
following counterterm \cite{Balasubramanian:1999re}:
\begin{equation}
I^{ct}_{g}=-\frac{1}{\kappa}\int_{\partial\mathcal{M}}{d^{3}x\sqrt
{-h}\biggl{(}\frac{2}{l}+\frac{\mathcal{R}l}{2}\biggr{)}}%
\end{equation}
Here, $h_{ab}$ is the induced metric on the boundary and $\mathcal{R}$ is its
Ricci scalar.

In the presence of the scalar field, this counterterm is not sufficient to
cancel the divergences in the action. For this case, an additional boundary
term that depends on the scalar is needed, namely $I_{\phi}$. We are going to
study the variational principle of the following action:
\begin{equation}
I=\int{d^{4}x\sqrt{-g}\biggl{(}\frac{R}{2\kappa}-\frac{(\partial\phi)^{2}}%
{2}-V(\phi)\biggr{)}}+\frac{1}{\kappa}\int_{\partial\mathcal{M}}{d^{3}%
x\sqrt{-h}K}-\frac{1}{\kappa}\int_{\partial\mathcal{M}}{d^{3}x\sqrt
{-h}\biggl{(}\frac{2}{l}+\frac{\mathcal{R}l}{2}\biggr{)}} + I_{\phi}^{ct}
\label{Icomplete}%
\end{equation}
for a scalar field with the conformal mass $m=-2l^{-2}$. In some
previous work (for example, see \cite{Lu:2013ura, Lu:2014maa}), the
following counterterm that produces a finite action for the
non-logarithmic branch was proposed:
\begin{equation}
\frac{1}{6\kappa}\int_{\partial\mathcal{M}}{d^{3}x\sqrt
{-h}\biggl{(}\phi n^{\nu}\partial_{\nu}\phi-\frac{\phi^{2}}{2l}\biggr{)}}
\label{phict}%
\end{equation}
However, it is problematic because it is not intrinsic to the boundary and
also, for mixed boundary conditions, the variational principle is not
satisfied. Instead, we propose new counterterms for both, the logarithmic and
non-logarithmic branches, so that the action is finite and there is a
well-posed variational principle. These intrinsic counterterms are constructed
to be compatible with the Hamiltonian method in the sense that the results
match for any boundary conditions.

Let us start with the non-logarithmic branch with the boundary term associated
to the scalar field given by
\begin{equation}
I_{\phi}^{ct}=-\int_{\partial\mathcal{M}}{d^{3}x\sqrt{-h}\biggl{[}\frac{\phi^{2}%
}{2l}+\frac{W(\alpha)}{l\alpha^{3}}\phi^{3}\biggr{]}} \label{ctphi}%
\end{equation}
Then, by using the boundary expansion of the metric and scalar field, the
variation of the action yields a boundary term evaluated at the cutoff $r$:
\begin{equation}
\delta I=\int{d^{3}x\sqrt{-h}\biggl{[}\frac{1}{r}\biggl{(}-\sqrt{g^{rr}}%
\phi^{^{\prime}}-\frac{\phi}{l}-\frac{3W(\alpha)\phi^{2}}{l\alpha^{3}%
}\biggr{)}\biggl{(}1+\frac{1}{r}\frac{d^{2} W(\alpha)}{d\alpha^{2}%
}\biggr{)}+\biggl{(}\frac{3W(\alpha)}{\alpha}-\beta\biggr{)}\frac{\phi^{3}%
}{l\alpha^{3}}\biggr{]}}\delta\alpha
\end{equation}
It is easy to show then that the variational principle is well defined when
the cutoff goes to infinity:
\begin{equation}
\lim_{r\to\infty}\delta I=0
\end{equation}
For the logarithmic branch we should work with the following counterterm for
the scalar field:
\begin{equation}
I_{\phi}^{ct}+\bar{I}_{\phi}^{ct}=-\int_{\partial\mathcal{M}}{d^{3}x\sqrt{-h}\biggl{[}\frac{\phi^{2}%
}{2l}+\frac{\phi^{3}}{l\alpha^{3}}\biggl{(}W-\frac{\alpha\gamma}%
{3}\biggr{)}-\frac{\phi^{3}C_{\gamma}}{3l}\ln\biggl{(}\frac{\phi}{\alpha
}\biggr{)}\biggr{]}}\label{ILOG}
\end{equation}
where $\bar{I}_{\phi}^{ct}$ is the counterterm necessary to provide a well possed acction principal in the case of the logartithmic branch. Using the asymptotic expansions of the metric and scalar field we also obtain
that the variational principle is well-defined for arbitrary boundary
conditions when the cut-off surface is send to infinity. As we shall show below, the same counterterms provide a finite on-shell action and the right free energy.

\subsection{Regularized action and free energy}

Evaluating the action leads to a formally divergent result. Now, we would like
to show that, indeed, all the divergences can be eliminated by using the
counterterms proposed in the previous section and so the action is finite. We
use the standard technique of Wick rotating the time direction $t=i\tau$.
Then, the temperature is related to the periodicity of the Euclidean time
$\tau$ ($\Delta\tau=\beta=1/T$) and the leading contribution to the free
energy is determined by evaluating the Euclidean action.

The action has four terms, the bulk part $I^{E}_{bulk}$, Gibbons-Hawking
surface term $I^{E}_{GH}$, and two boundary counterterms ($I_{g}^{ct}$,
$I^{ct}_{\phi}$): $I= I^{E}_{bulk} + I^{E}_{GH} + I_{g}^{ct} + I_{\phi}^{ct}$.
Let us compute these contributions for the non-logarithmic branch first.  

Since we are going to study the properties of a large family of exact hairy
black hole solutions, let us start with the following generic metric ansatz
\begin{equation}
ds^{2}=\Omega(x)\left[  -f(x)dt^{2}+\frac{\eta^{2}dx^{2}}{f(x)}+d\Sigma
_{k}^{2}\right]  \label{Ansatz}%
\end{equation}
Concrete expressions for the functions $\Omega(x)$ and $f(x)$ are presented in
Section \ref{exact3}.

The computations in the $(t,x,\Sigma)$ coordinate system (\ref{Ansatz}) are
related by a simple coordinate transformation to $(t,r,\Sigma)$ system
(\ref{Ansatz1}). In
what follows, $x_{b}$ and $r_{b}$ denote the boundary, and $x_{+}$
and $r_{+}$  the horizon. The on-shell Euclidean bulk action can be written as
\begin{equation}
I_{bulk}^{E}=\int_{0}^{1/T}{d\tau}\int_{x_{+}}^{x_{b}}{d^{3}x\sqrt{g^{E}}%
V}\left(  \phi\right)  =\frac{\sigma_{k}}{2\eta\kappa T}\frac{d(\Omega f)}%
{dx}\biggr{\vert}_{x_{+}}^{x_{b}}%
\end{equation}
where $\sigma_{k}$ is the area of $\Sigma_{k}$ (e.g., for $k=1$ $\sigma
_{1}=4\pi)$ and $g^{E}$ is the metric on the Euclidean section. The two
coordinate systems $(t,x,\Sigma_{k})$ and $(t,r,\Sigma_{k})$ are related by
\begin{equation}
\Omega(x)\rightarrow S(r);\qquad f(x)\rightarrow\frac{N(r)}{S(r)};\qquad
dx\rightarrow\frac{\sqrt{NH}}{\eta S}dr\label{Changeco}%
\end{equation}
and so we can rewrite the bulk integral result in the coordinates
$(t,r,\Sigma_{k})$ as
\begin{equation}
I_{bulk}^{E}=\frac{\sigma_{k}}{2\kappa T}\frac{S}{\sqrt{NH}}\frac{dN}%
{dr}\biggr{\vert}_{r_{+}}^{r_{b}}%
\end{equation}
Let us now compute the Gibbons-Hawking term. Consider a timelike hypersurface
$x=x_{0}$, then the induced metric $h^{\mu\nu}=g^{\mu\nu}-n^{\mu}n^{\nu}$,
normal, extrinsic curvature, and its trace $K=h^{\mu\nu}K_{\mu\nu}$ are
\begin{equation}
ds^{2}=h_{ab}dx^{a}dx^{b}=\Omega(x_{0})\biggl{[}-f(x_{0})dt^{2}+d\Sigma
_{k}\biggr{]}
\end{equation}%
\begin{equation}
n_{a}=\frac{\delta_{a}^{x}}{\sqrt{g^{xx}}}\biggr{\vert}_{x=x_{0}};\qquad
K_{ab}=\frac{\sqrt{g^{xx}}}{2}\partial_{x}g_{ab}\biggr{\vert}_{x=x_{0}};\qquad
K=\frac{1}{2\eta}\biggl{(}\frac{f}{\Omega}\biggr{)}^{1/2}\biggl{[}\frac
{(\Omega f)^{^{\prime}}}{\Omega f}+\frac{2\Omega^{^{\prime}}}{\Omega
}\biggr{]}\biggr{\vert}_{x_{0}}\label{nkk}%
\end{equation}
and using the transformation equations (\ref{Changeco}) the contribution of
this term can be rewritten as
\begin{equation}
I_{GH}^{E}=-\frac{\sigma_{k}}{\kappa T}\frac{\Omega f}{2\eta}\biggl{[}\frac
{(\Omega f)^{^{\prime}}}{\Omega f}+\frac{2\Omega^{^{\prime}}}{\Omega
}\biggr{]}\biggr{\vert}_{x_{b}}=-\frac{\sigma_{k}}{2T\kappa}\biggl{(}\frac
{S}{\sqrt{NH}}\frac{dN}{dr}+\frac{2N}{\sqrt{NH}}\frac{dS}{dr}%
\biggr{)}\biggr{\vert}_{r_{b}}%
\end{equation}
The contribution from the gravitational counterterm is
\begin{equation}
I_{g}^{ct}=\frac{2\sigma_{k}}{\kappa Tl}\biggl{(}\Omega^{3/2}f^{1/2}%
+\frac{l^{2}k}{2}f^{1/2}\Omega^{1/2}\biggr{)}\biggr{\vert}_{x_{b}}%
=\frac{2\sigma_{k}}{\kappa Tl}S\sqrt{N}\biggl{(}1+\frac{l^{2}k}{2S}%
\biggr{)}\biggr{\vert}_{r_{b}}%
\end{equation}

Using the general formula for the temperature
\begin{equation}
T=\frac{N^{^{\prime}}}{4\pi\sqrt{NH}}\biggr{\vert}_{r_{+}}%
\end{equation}
one can write the sum of these three contributions in the total action as
\begin{equation}
I_{bulk}^{E}+I_{GH}^{E}+I^{ct}_{g}=-\frac{1}{T}\biggl{[}\frac{\sigma
_{k}S(r_{+})T}{4G}\biggr{]}-\frac{\sigma_{k}}{2\kappa T}\biggl{[}\frac
{2N}{\sqrt{NH}}\frac{dS}{dr}-\frac{4}{l}S\sqrt{N}\biggl{(}1+\frac{l^{2}k}%
{2S}\biggr{)}\biggr{]}\biggr{\vert}_{r_{b}} \label{III}%
\end{equation}
which, for a scalar field with the conformal mass $m^{2}=-2l^{-2}$ as in
(\ref{phi}) and with the metric fall-off (\ref{gtt}) and (\ref{grr}), becomes
\begin{equation}
I_{bulk}^{E}+I_{GH}^{E}+I^{ct}_{g}=-\frac{\mathcal{A}}{4G }-\frac{\sigma_{k}}{T}\biggl{(}-\frac{\mu}{\kappa}%
+\frac{4\alpha\beta}{3l^{2}}+\frac{r\alpha^{2}}{2l^{2}}%
\biggr{)}\biggr{\vert}_{r_{b}}%
\end{equation}
Here, $\mathcal{A}=\sigma_{k}S(r_{+})$ is the horizon area.

It is clear now that the gravitational counterterm is not sufficient to remove
the divergences at the boundary $r_{b}\rightarrow\infty$, but this new linear
divergence can be regularized with the following counterterm that depends on
the scalar field:
\begin{equation}
I_{\phi}^{ct}=\int_{\mathcal{\partial M}}{d^{3}x\sqrt{h^{E}}\biggl{[}\frac
{\phi^{2}}{2l}+\frac{W(\alpha)}{l\alpha^{3}}\phi^{3}\biggr{]}}=\frac
{\sigma_{k}}{T}\biggl{(}\frac{W}{l^{2}}+\frac{\alpha\beta}{l^{2}}%
+\frac{r\alpha^{2}}{2l^{2}}\biggr{)}\biggr{\vert}_{r_{\infty}} \label{Iphi}%
\end{equation}
The renormalized Euclidean action can be rewritten then using $\beta
=dW/d\alpha$ as
\begin{equation}
I^{E}=I_{bulk}^{E}+I_{GH}^{E}+I_{g}^{ct}+I_{\phi}^{ct} =-\frac{\mathcal{A}%
}{4G} + \frac{\sigma_{k}}{T}\biggl{[}\frac{\mu}{\kappa}+\frac{1}{l^{2}%
}\biggl{(}W-\frac{\alpha}{3}\frac{dW}{d\alpha}\biggr{)}\biggr{]}
\label{Fnonlog}%
\end{equation}
and so the free energy becomes
\begin{equation}
F=I^{E}T=M - TS
\end{equation}
The thermodynamic relations will provide the same mass and entropy for the
black holes:
\begin{equation}
M=-T^{2}\frac{\partial I^{E}}{\partial T}=\sigma_{k}\biggl{[}\frac{\mu}%
{\kappa}+\frac{1}{l^{2}}\biggl{(}W-\frac{\alpha}{3}\frac{dW}{d\alpha
}\biggr{)}\biggr{]}
\end{equation}
and
\begin{equation}
S=-\frac{\partial(I^{E}T)}{\partial T}=\frac{\mathcal{A}}{4G}%
\end{equation}
A similar computation can be done for the logarithmic branch. We work again
with a scalar field with the conformal mass $m^{2}=-2l^{-2}$ with a fall-off
(\ref{philog}) for which the metric fall-off is (\ref{grrlog}). If we work
with the counterterm (\ref{ctphi}), we obtain
\begin{equation}
I_{bulk}^{E}+I_{GH}^{E}+I^{ct}_{g}+I_{\phi}^{ct}=-\frac{\mathcal{A}}{4G}
+\frac{\sigma_{k}}{T}\biggl{\lbrace}\frac{\mu}{\kappa}+\frac{1}{l^{2}%
}\biggl{[}W(\alpha)-\frac{\alpha}{3}\frac{dW}{d\alpha}+\frac{2\alpha\gamma}%
{9}-\frac{\alpha\gamma}{3}\ln{r}\biggr{]}\biggr{\rbrace}
\end{equation}
and we see that there is still a logarithmic divergence. Therefore, one
consider a new contribution from the scalar field that will also cancel that
divergence (\ref{ILOG}):
\begin{equation}
\bar{I}^{\,\,ct}_{\phi}=\int_{\partial\mathcal{M}}{d^{3}x\sqrt{h^{E}%
}\biggl{\lbrace}\frac{\phi^{3}\gamma}{3\alpha^{2}l}\biggl{[}\ln\biggl{(}\frac
{\alpha}{\phi}\biggr{)}-1\biggr{]}\biggr{\rbrace}}=\frac{\sigma_{k}}%
{T}\biggl{[}-\frac{\alpha\gamma}{3l^{2}}+\frac{\alpha\gamma\ln{r}}{3l^{2}%
}+O(r^{-1}\ln{r})\biggr{]} \label{newlogct}%
\end{equation}
We also obtain a finite action for the logarithmic branch
\begin{equation}
I^{E} =I_{bulk}^{E}+I_{GH}^{E}+I^{ct}_{g}+I_{\phi}^{ct}+\bar{I}^{\,\,ct}%
_{\phi} = -\frac{\mathcal{A}}{4G}+\frac{\sigma_{k}}{T}\biggl{[}\frac{\mu
}{\kappa}+\frac{1}{l^{2}}\biggl{(}W-\frac{\alpha}{3}\frac{dW}{d\alpha}%
-\frac{\alpha\gamma}{9}\biggr{)}\biggr{]} \label{Flog}%
\end{equation}
where $\gamma=C_{\gamma}\alpha^{2}$ and $C_{\gamma}=-3l^{2}\lambda$.

The thermodynamic relations provide the right results:
\begin{equation}
M=-T^{2}\frac{\partial I^{E}}{\partial T}=\sigma_{k}\biggl{[}\frac{\mu}{\kappa}+\frac
{1}{l^{2}}\biggl{(}W-\frac{\alpha}{3}\frac{dW}{d\alpha}-\frac{\alpha\gamma}%
{9}\biggr{)}\biggr{]}
\end{equation}
and
\begin{equation}
S=-\frac{\partial(I^{E}T)}{\partial T}=\frac{\mathcal{A}}{4G}%
\end{equation}
The conformal symmetry of the boundary is preserved when $W(\alpha)=\alpha
^{3}(C+l^{2}\lambda\ln{\alpha})$.

\section{Regularized Brown-York stress tensor}

Within the AdS/CFT duality, the AdS graviton couples to the stress-energy
tensor of the CFT \cite{Gubser:1997yh,Gubser:1997se}:
\begin{equation}
\int_{\partial\mathcal{M}} d^{3}x\, h^{ab}\,T_{ab}\
\end{equation}
Then, from a holographic point of view, the Brown-York stress tensor is
interpreted as the stress-energy tensor of the dual field theory. In this
section we work in the coordinates $(t,r,\Sigma)_{k}$ for which the metric was
given in (\ref{Ansatz1}). For the bulk geometry we use the foliation with the
surfaces $r=R=constant$ and the induced metric is
\begin{equation}
ds^{2}=h_{ab}dx^{a}dx^{b}=-N(R)dt^{2}+S(R)d\Sigma_{k}^{2} \label{k1induced}%
\end{equation}
The Brown-York (quasilocal) stress tensor is defined as \cite{Brown:1992br}
\begin{equation}
\tau^{ab}\equiv\frac{2}{\sqrt{-h}}\frac{\delta I}{\delta h_{ab}}%
\end{equation}
where $I$ is the total action including the counterterms.

Since the metric where the dual field lives is related to the boundary metric
by a conformal factor, it is very important to emphasize that the CFT stress
tensor is also related to the Brown-York stress tensor up to a conformal
factor. As an warm up exercise, let us describe this method for the
$4$-dimensional Schwarzschild-AdS black hole --- we are going to follow the
analysis of \cite{Myers:1999psa}.

The black hole metric is
\begin{equation}
ds^{2}=-\biggl{(}1-\frac{m}{r}+\frac{r^{2}}{l^{2}}\biggr{)}dt^{2}%
+\biggl{(}1-\frac{m}{r}+\frac{r^{2}}{l^{2}}\biggr{)}^{-1}dr^{2}+r^{2}%
d\Omega^{2}%
\end{equation}
and if we consider the foliation $r=R$ the induced metric $h_{ab}$ of any
`slice' is
\begin{equation}
ds^{2}=-\biggl{(}1-\frac{m}{R}+\frac{R^{2}}{l^{2}}\biggr{)}dt^{2}+R^{2}%
d\Omega^{2}%
\end{equation}


As we have pointed out before, the boundary metric is
\begin{equation}
ds^{2}_{boundary}=\frac{R^{2}}{l^{2}}(-dt^{2}+l^{2}d\Omega^{2})
\end{equation}
but the background metric where the dual quantum field theory lives is
$\gamma_{ab}$ defined as
\begin{equation}
ds_{dual}^{2}=\gamma_{ab}dx^{a}dx^{b}=-dt^{2}+l^{2}d\Omega^{2}%
\end{equation}
The metric $\gamma_{ab}$ is not dynamical and it is related by a conformal
factor to the boundary metric. The corresponding dual stress tensor is
\begin{equation}
\langle\tau_{ab}^{dual}\rangle=\lim_{R\rightarrow\infty}\frac{R}{l}\tau
_{ab}=\frac{m}{16\pi G l^{2}}[3\delta_{a}^{0}\delta_{b}^{0}+\gamma_{ab}]
\end{equation}
Written in this way \cite{Myers:1999psa} it has the form of a thermal gas of a
massless particles and, as expected, its trace vanishes $\langle\tau
^{dual}\rangle=\langle\tau_{ab}^{dual}\rangle\gamma^{ab}=0$.

A similar procedure can be used for the hairy black holes, but one should add
the boundary counterterms related to the scalar field. In the case of the
non-logarithmic branch, the complete action is (\ref{Icomplete}) and the
scalar counterterm was given in (\ref{phict}), where $G_{ab}$ is the Einstein
tensor for the foliation (\ref{k1induced}) given by $G_{ab}=\delta_{a}%
^{t}\delta_{b}^{t}Nk/S$. The regularized stress tensor is
\begin{equation}
\tau_{ab}=-\frac{1}{\kappa}\biggl{(}K_{ab}-h_{ab}K+\frac{2}{l}h_{ab}%
-lG_{ab}\biggr{)}-\frac{h_{ab}}{l}\biggl{[}\frac{\phi^{2}}{2}+\frac{W(\alpha
)}{\alpha^{3}}\phi^{3}\biggr{]}\label{BY1}%
\end{equation}
Thus, the stress tensor components are
\begin{align}
\tau_{tt} &  =\frac{l}{R}\biggl{[}\frac{\mu}{8\pi G l^{2}}+\frac{1}{l^{4}%
}\biggl{(}W-\frac{\alpha\beta}{3}\biggr{)}\biggr{]}+O(R^{-2})\\
\tau_{\theta\theta} &  =\frac{l}{R}\biggl{[}\frac{\mu}{16\pi G}-\frac
{1}{l^{2}}\biggl{(}W-\frac{\alpha\beta}{3}\biggr{)}\biggr{]}+O(R^{-2}%
)\nonumber\\
\tau_{\phi\phi} &  =\frac{l\sin^{2}{\theta}}{R}\biggl{[}\frac{\mu}{16\pi
G}-\frac{1}{l^{2}}\biggl{(}W-\frac{\alpha\beta}{3}%
\biggr{)}\biggr{]}+O(R^{-2})\nonumber
\end{align}
The stress tensor of the dual field theory can be put in a similar form as for
the Schwarzschild-AdS black hole:
\begin{equation}
\langle\tau_{ab}^{dual}\rangle=\frac{3\mu}{16\pi G l^{2}}\delta_{a}%
^{0}\delta_{b}^{0}+\frac{\gamma_{ab}}{l^{2}}\biggl{[}\frac{\mu}{16\pi G %
}-\frac{1}{l^{2}}\left(  W(\alpha)-\frac{\alpha\beta}{3}\right)  \biggr{]}
\end{equation}
The trace can be easily computed and we get
\begin{equation}
\langle\tau^{dual}\rangle=-\frac{3}{l^{4}}\biggl{[}W(\alpha)-\frac{\alpha
\beta}{3}\biggr{]}
\end{equation}
Unlike the Schwarzschild-AdS black hole, for the hairy black holes there are
two different types of boundary conditions, namely that preserve or not the
conformal symmetry. As expected, when the conformal symmetry is preserved
$W=C\alpha^{3}$ the trace of the dual stress tensor vanishes $\langle
\tau^{dual}\rangle=0$.

A similar, but more complicated, procedure can be applied for the logarithmic
branch. The action (\ref{Icomplete}) has a new contribution (\ref{newlogct})
that cancel the logarithmic divergence, and the new regularized quasilocal
stress tensor is
\begin{equation}
\tau_{ab}=-\frac{1}{\kappa}\biggl{(}K_{ab}-h_{ab}K+\frac{2}{l}h_{ab}%
-lG_{ab}\biggr{)}-\frac{h_{ab}}{l}\biggl{[}\frac{\phi^{2}}{2}+\frac{\phi^{3}%
}{\alpha^{3}}\biggl{(}W-\frac{\alpha\gamma}{3}\biggr{)}+\frac{\phi^{3}\gamma
}{3\alpha^{2}}\ln{\biggl{(}\frac{\alpha}{\phi}\biggr{)}}\biggr{]}\label{BY2}%
\end{equation}
with the following components
\begin{align}
\tau_{tt} &  =\frac{l}{R}\biggl{[}\frac{\mu}{8\pi G l^{2}}+\frac{1}{l^{4}%
}\biggl{(}W-\frac{\alpha\beta}{3}-\frac{\alpha\gamma}{9}%
\biggr{)}\biggr{]}+O\biggl{[}\frac{(\ln{R})^{3}}{R^{2}}\biggr{]}\\
\tau_{\theta\theta} &  =\frac{l}{R}\biggl{[}\frac{\mu}{16\pi G }-\frac
{1}{l^{2}}\biggl{(}W-\frac{\alpha\beta}{3}-\frac{\alpha\gamma}{9}%
\biggr{)}\biggr{]}+O\biggl{[}\frac{(\ln{R})^{3}}{R^{2}}\biggr{]}\nonumber\\
\tau_{\phi\phi} &  =\frac{l\sin^{2}{\theta}}{R}\biggl{[}\frac{\mu}{16\pi
G }-\frac{1}{l^{2}}\biggl{(}W-\frac{\alpha\beta}{3}-\frac{\alpha\gamma}%
{9}\biggr{)}\biggr{]}+O\biggl{[}\frac{(\ln{R})^{3}}{R^{2}}\biggr{]}\nonumber
\end{align}
and so the stress tensor of the dual field theory becomes
\begin{equation}
\langle\tau_{ab}^{dual}\rangle=\frac{3\mu}{16\pi G l^{2}}\delta_{a}%
^{0}\delta_{b}^{0}+\frac{\gamma_{ab}}{l^{2}}\biggl{[}\frac{\mu}{16\pi G %
}-\frac{1}{l^{2}}\left(  W(\alpha)-\frac{\alpha\beta}{3}-\frac{\alpha\gamma
}{9}\right)  \biggr{]}
\end{equation}
Its trace is
\begin{equation}
\langle\tau^{dual}\rangle=-\frac{3}{l^{4}}\biggl{(}W-\frac{\alpha\beta}%
{3}-\frac{\alpha\gamma}{9}\biggr{)}
\end{equation}
and, as expected, it vanishes for the boundary conditions that preserve the
conformal symmetry:
\begin{equation}
\langle\tau^{dual}\rangle=0\Rightarrow\gamma=-3l^{2}\lambda\alpha^{2};\qquad
W(\alpha)=\alpha^{3}[C+l^{2}\lambda\ln{\alpha}]
\end{equation}

\section{Hamiltonian mass and holographic mass}

In AdS spacetime there exist different methods of computing the gravitational
mass and a comparison between them is going to be useful --- for Dirichlet
boundary conditions this was done in great detail in \cite{Hollands:2005wt}
and up to some ambiguities related to constant boundary terms, the Hamiltonian
formalism, AMD mass, and holographic method produce the same result. Though,
as was pointed out in \cite{Anabalon:2014fla}, when the conformal symmetry is
broken in the boundary the AMD mass is not the correct physical mass and one
should compute the Hamiltonian mass of the system. In this section we provide
details of computing the Hamiltonian mass and show that it matches the
holographic mass even when the conformal symmetry in the boundary is broken.

\subsection{Hamiltonian formalism}

We consider the Regge-Teitelboim approach \cite{Regge:1974zd} to compute the
mass of static scalar hairy asymptotically locally AdS spacetimes. A summary
of this method is provided below. We are considering the action (\ref{action})
for which the Hamiltonian constraints $\mathcal{H}_{\bot}$ and $\mathcal{H}%
_{i}$, with $i=1,2,3$, contain contributions from the gravitational term and
from the matter sector that in this case corresponds to a minimally coupled
scalar field with a self-interaction potential $V(\phi)$. These constraints
are functions of the canonical variables: the three-dimensional metric
$g_{ij}$ and the scalar field $\phi$, and their corresponding conjugate
momenta $\pi^{i j}$ and $\pi_{\phi}$. The Hamiltonian constraints are given
by
\begin{align}
{\mathcal{H}}_{\bot}  &  =\frac{2\kappa}{\sqrt{g }}\left[  \pi_{ij}\pi
^{ij}-\frac{1}{2}\left(  \pi^{i}{}_{i}\right)  ^{2}\right]  - \frac{1}%
{2\kappa}\sqrt{g }\left.  ^{(3)}R \right. \nonumber\\
&  + \frac{1}{2}\left(  \frac{\pi_{\phi}{}^{2}}{\sqrt{g }}+\sqrt{g }g
^{ij}\phi,_{i}\phi,_{j}\right)  +\sqrt{g }V\left(  \phi\right) \label{hperp}\\
{\mathcal{H}}_{i}  &  =-2\pi_{i}^{j}{}_{\mid j} + \pi_{\phi}\phi,_{i}
\label{hi}%
\end{align}
The three-dimensional metric $g_{ij}$ can be recognized from the line element
written in its ADM form
\begin{equation}
\label{ADMform}ds^{2}=-(N^{\perp})^{2}dt^{2}+g _{ij}\left(  dx^{i}%
+N^{i}dt\right)  \left(  dx^{j}+N^{j}dt\right)
\end{equation}
and, $g$, $^{(3)}R$ and vertical bar $\mid$ denote the determinant, the scalar
curvature, and the covariant derivative associated to the spatial metric, respectively.

The canonical generator of an asymptotic symmetry defined by the vector
$\xi=(\xi^{\perp},\xi^{i})$ is a linear combination of the constraints
$\mathcal{H}_{\perp}, \mathcal{H}_{i}$ plus a surface term $Q[\xi]$
\begin{equation}
\label{congen}H[\xi]=\int_{\partial\mathcal{M}} d^{3} x \left(  \xi^{\perp}
\mathcal{H}_{\perp}+\xi^{i}\mathcal{H}_{i}\right)  +Q[\xi]
\end{equation}
$Q[\xi]$ is chosen in order to cancel out the surface terms coming from the
variation of the generator with respect to the canonical variables. In this
way, the generator $H[\xi]$ possesses well-defined functional derivatives
\cite{Regge:1974zd}. The general form of $Q[\xi]$ for the generator
(\ref{congen}) \cite{Henneaux:2006hk} is given by
\begin{align}
\label{de}\delta Q[\xi]  &  =\oint d^{2}S_{l}\left[  \frac{G^{ijkl}}{2 \kappa
}(\xi^{\bot}\delta g_{ij}{}_{\mid k}-{\xi^{\bot}}_{,k} \delta g_{ij})+2
\xi_{k}\delta\pi^{kl}\right. \nonumber\\
&  \left.  +(2 \xi^{k}\pi^{jl}-\xi^{l}\pi^{jk}) \delta g_{jk}- (\sqrt{g}
\xi^{\bot}g^{lj}\phi,_{j}+\xi^{l}\pi_{\phi})\delta\phi\right]
\end{align}
where
\begin{equation}
G^{ijkl}\equiv\frac{1}{2}\sqrt{g}(g^{ik}g^{jl}+g^{il}g^{jk}-2g^{ij}g^{kl})
\end{equation}
The normal and tangential components of the allowed deformation $(\xi^{\perp
},\xi^{i})$ are related with the spacetime components $(\xi^{\perp},{}%
^{(3)}\xi^{i})$ in the following way
\begin{equation}
\xi^{\perp} =N^{\perp} \xi^{t}, \quad\xi^{i} ={}^{(3)}\xi^{i}+N^{i}\xi^{t}%
\end{equation}

The following step is to note that the Hamiltonian generator (\ref{congen})
reduces to the surface term $Q[\xi]$ when the constraints hold. Thus, the
value of the generators --- the conserved charges associated the the
asymptotic symmetries --- are just given by $Q[\xi]$. Since the charges are
defined by a surface term at the boundary, they need just the behaviour of the
canonical variables and symmetries close to the boundary. i.e. their
asymptotic behavior. Thus, the charges obtained from the Hamiltonian method
are appropriate for a holographic interpretation. Additionally, one can remark
that the canonical generators provide the charges for all the solutions
sharing the same asymptotic behaviour.

We focus now in the static case. By definition there is a timelike Killing
vector $\partial_{t}$, and the corresponding conserved charge associated with
this symmetry ---time translation--- is from first principles the mass $M$. In
the static case all the momenta vanish and the expression (\ref{de}),
evaluated for $\xi= \partial_{t}$, reduces to
\begin{equation}
\label{qt}\delta M\equiv\delta Q[\partial_{t}]=\oint d^{2}S_{l} \left[
\frac{G^{ijkl}}{2 \kappa}(\xi^{\bot}\delta g_{ij}{}_{\mid k}-{\xi^{\bot}}_{,k}
\delta g_{ij})- \sqrt{g} \xi^{\bot}g^{lj}\phi,_{j}\delta\phi\right]
\end{equation}
We note an explicit contribution of the scalar field in the mass that, in
general, yields a non-vanishing amount. In order to achieve a better
understanding of this contribution, it is convenient to separate it from the
usual gravitational contribution by writing $\delta M$ as
\begin{equation}
\delta M=\delta M_{G}+\delta M_{\phi}%
\end{equation}
where
\begin{equation}
\label{eq:Q_G}\delta M_{G}=\oint d^{2}S_{l} \frac{G^{ijkl}}{2 \kappa}%
(\xi^{\bot}\delta g_{ij}{}_{\mid k}-{\xi^{\bot}}_{,k} \delta g_{ij})
\end{equation}
and
\begin{equation}
\label{eq:Q_phi}\delta M_{\phi}=-\oint d^{2}S_{l} \sqrt{g} \xi^{\bot}%
g^{lj}\phi,_{j}\delta\phi
\end{equation}

As mentioned before, the variation of the mass, given by surface integral
(\ref{qt}), needs just the asymptotic behavior of the canonical variables and
symmetries. However, this variation usually requires more information to be
integrated, and boundary conditions must be imposed. The necessity of boundary
conditions is expected from physical grounds, since the mass of a system is
well defined after imposing suitable boundary conditions. The effect of a slow
fall-off scalar field on the mass of asymptotically hairy spacetimes have been
studied in \cite{Henneaux:2002wm, Henneaux:2004zi,Henneaux:2006hk} using the
Hamiltonian formalism described above. Other approaches and methods
\cite{Barnich:2002pi,Gegenberg:2003jr,Hertog:2004dr,Banados:2005hm,Amsel:2006uf}
have confirmed this effect.

One step further was made in \cite{Anabalon:2014fla} where the computation of
the mass of these hairy configurations was done by considering the additional
information provided by the remaining field equations. For this work, we focus
on the analysis on the class of potentials having a mass term corresponding to
the conformal mass $m^{2}=-2l^{-2}$ in four dimensions. For completeness and
because the boundary conditions and the way the divergences cancel in the
construction of the Hamiltonian mass provide helpful intuition for
constructing the counterterms, in the next subsection we present the details
of the analysis of \cite{Anabalon:2014fla}.

\subsection{Non-logarithmic and logarithmic branches for $m^{2}=-2/l^{2}$}

Expanding the potential as a power series around $\phi=0$, it was shown
\cite{Henneaux:2006hk} the absence of logarithmic branches in the
asymptotically behavior of the metric and scalar field, provided the series
does not contain a cubic term. This set of asymptotic conditions accommodates
exact scalar black hole solutions
\cite{Martinez:2004nb,Anabalon:2012ta,Acena:2013jya, Anabalon:2013eaa} whose
asymptotic behavior belong the chosen one. The fall-off of the scalar field
and metric at infinity was obtained in Section $2$.

Now, we evaluate the general expressions (\ref{eq:Q_G}) and (\ref{eq:Q_phi})
for static configurations, using the above asymptotic conditions. We consider
a boundary located at $r=\infty$. Integrating the `angular coordinates', we
obtain the gravitational contribution
\begin{equation}
\delta M_{G}=\frac{\sigma_{k}}{\kappa}[r\delta a+l\delta b+O(1/r)]
\label{eq:delta_mg}%
\end{equation}
and the contribution from the scalar field
\begin{equation}
\delta M_{\phi}=\frac{\sigma_{k}}{l^{2}}[r\alpha\delta\alpha+\alpha\delta
\beta+2\beta\delta\alpha+O(1/r)] \label{eq:delta_mphi}%
\end{equation}
By adding both contributions we have the variation of the mass
\begin{equation}
\delta M=\frac{\sigma_{k}}{\kappa l^{2}}[r(l^{2}\delta a+\kappa\alpha
\delta\alpha)+l^{3}\delta b+\kappa(\alpha\delta\beta+2\beta\delta
\alpha)+O(1/r)] \label{varmass}%
\end{equation}
It is important to remind that this expression for $\delta M$ is meaningful
only in the case of vanishing constraints. In the static case, there is a
single nontrivial constraint, $H_{\perp}=0$, which for the asymptotic
conditions displayed above yields
\begin{equation}
\frac{k+a}{\kappa}+\frac{\alpha^{2}}{2l^{2}}=0 \label{a}%
\end{equation}
The linear divergent piece in (\ref{varmass}) is removed by replacing
(\ref{a}) into (\ref{varmass}). Then, the asymptotic variation of the mass
becomes finite
\begin{equation}
\delta M=\frac{\sigma_{k}}{\kappa l^{2}}[l^{3}\delta b+\kappa(\alpha
\delta\beta+2\beta\delta\alpha)] \label{varmassfin}%
\end{equation}

In order to integrate the variations in (\ref{varmassfin}) boundary conditions
on the scalar field are necessary. In particular, the integration of
(\ref{varmassfin}) requires a functional relation between $\alpha$ and $\beta
$. If we define $\beta=dW(\alpha)/d\alpha$, the mass of the spacetime is given
by
\begin{equation}
\label{eq:mass1}M=\sigma_{k}\left[  \frac{l b}{\kappa}+\frac{1}{l^{2}}\left(
\alpha\frac{dW(\alpha)}{d\alpha}+W(\alpha)\right)  \right]
\end{equation}
We note that the mass in (\ref{eq:mass1}) is defined up to a constant without
variation. This constant is set to be zero in order to fix a vanishing mass
for the locally AdS spacetime because in four dimensions there is no Casimir energy.

To obtain the logarithmic branch, it is necessary to use the self-interaction
potential (\ref{vphilog}) so that the fall-off of the scalar field to be
considered is (\ref{philog}). The Hamiltonian constraint $H_{\perp}=0$ is
satisfied if (\ref{a}) and
\begin{equation}
\frac{lc}{\kappa}-4\alpha^{3}\lambda=0 \label{d}%
\end{equation}
are fulfilled.

Now, we evaluate (\ref{eq:Q_G}) and (\ref{eq:Q_phi}). In this case we find
\begin{equation}
\label{mgl}\delta M_{G}=\biggr{\lbrace} \frac{l\delta b}{\kappa}+\frac{\delta
a}{\kappa} r+\frac{l\delta c}{\kappa} \ln(r)+O\left(  \frac{\ln(r)^{2}}%
{r}\right)  \biggl{\rbrace} \sigma_{k}%
\end{equation}
and
\begin{align}
\label{mpl}\delta M_{\phi}  &  =\left[  \frac{\alpha\delta\beta+2\beta
\delta\alpha+3\alpha^{2}l^{2}\lambda\delta\alpha}{l^{2}}+r\frac{\alpha
\delta\alpha}{l^{2}}\right. \nonumber\\
&  \left.  -12\lambda\alpha^{2}\delta\alpha\ln(r)+O\left(  \frac{\ln(r)^{2}%
}{r}\right)  \right]  \sigma_{k}%
\end{align}
Both contributions contain linear and logarithmic divergences. Adding
(\ref{mgl}) and (\ref{mpl}), the linear divergence cancels out by virtue of
(\ref{a}) and the logarithmic divergence vanishes by considering (\ref{d}).
Thus, we obtain a finite expression for the variation of the mass,
\begin{equation}
\delta M=\left[  \frac{l\delta b}{\kappa}+\frac{\alpha\delta\beta+2\beta
\delta\alpha+3\alpha^{2}l^{2}\lambda\delta\alpha}{l^{2}}\right]  \sigma_{k}%
\end{equation}
Again, we need a boundary condition, a functional relation between $\alpha$
and $\beta$, in order to integrate $\delta M$. We consider the general
relation $\beta=\frac{dW}{d\alpha}$, so that the Hamiltonian mass is given by
\begin{equation}
M=\left[  \frac{lb}{\kappa}+\frac{1}{l^{2}}\left(  \alpha\frac{dW}{d\alpha
}+W(\alpha)+\alpha^{3}l^{2}\lambda\right)  \right]  \sigma_{k}%
\end{equation}

The mass can be related with the first subleading term of $g_{tt}$ by using
(\ref{abglog}). Thus the mass can be written as
\begin{equation}
M=\left[  \frac{\mu}{\kappa}+\frac{1}{l^{2}}\left(  W(\alpha)-\frac{1}%
{3}\alpha\frac{dW}{d\alpha}+\frac{1}{3}\alpha^{3}l^{2}\lambda\right)  \right]
\sigma_{k}%
\end{equation}
Therefore, the expression $M=\mu\sigma_{k}\kappa^{-1}$ is obtained only for
$\alpha=0$ or
\begin{equation}
W(\alpha)=\alpha^{3}\left[  C+l^{2}\lambda\ln(\alpha)\right]
\end{equation}
which correspond to AdS invariant boundary conditions \cite{Henneaux:2006hk}.

\subsection{Holographic mass matches Hamiltonian mass}

Armed with the Brown-York formalism supplemented with counterterms, one can
obtain the energy of a hairy black holes. The boundary metric can be written,
at least locally, in ADM-like form. Provided the boundary geometry has an
isometry generated by the Killing vector $\xi^{a}=(\partial_{t})^{a}$ , the
energy is, as usual, the conserved charge.

Concretely, we are going to use the coordinates $(t,r,\Sigma_{k})$ with the
metric (\ref{Ansatz1}) and the foliation (\ref{k1induced}) parametrized as
\begin{equation}
d\Sigma_{k}^{2}=\frac{dy^{2}}{1-ky^{2}}+(1-ky^{2})d\phi^{2}%
\end{equation}
The energy
\begin{equation}
E=\int{d\sigma^{i}\tau_{ij}\xi^{j}}=\int{dyd\phi Su^{i}\tau_{ij}\xi^{j}}%
\end{equation}
is associated with the surface $t=constant$, for which the induced metric is
\begin{equation}
ds^{2}=\sigma_{ij}dx^{i}dx^{j}=Sd\Sigma_{k}^{2}%
\end{equation}
with the normal vector $u^{a}=N^{-1/2}(\partial_{t})^{a}$.

For the non-logarithmic branch, using the quasilocal stress tensor
(\ref{BY1}), one obtains
\begin{equation}
E=\sigma_{k}\biggl{[}\frac{\mu}{\kappa}+\frac{1}{l^{2}}\biggl{(}W-\frac
{\alpha}{3}\frac{dW}{d\alpha}\biggr{)}\biggr{]}
\end{equation}
With a similar computation for the logarithmic branch, but with the quasilocal
stress tensor (\ref{BY2}), we obtain the following energy of the hairy black
hole:
\begin{equation}
E=\sigma_{k}\biggl{[}\frac{\mu}{\kappa}+\frac{1}{l^{2}}\biggl{(}W-\frac{1}%
{3}\alpha\frac{dW}{d\alpha}-\frac{\alpha\gamma}{9}\biggr{)}\biggr{]} =
\sigma_{k}\biggl{[}\frac{\mu}{\kappa}+\frac{1}{l^{2}}\biggl{(}W-\frac{1}%
{3}\alpha\frac{dW}{d\alpha}-\frac{\alpha^{3}C_{\gamma}}{9}\biggr{)}\biggr{]}
\end{equation}
This shows perfect agreement with the Hamiltonian mass even if the conformal
symmetry is broken in the boundary --- with both methods it is possible to
obtain a finite energy in this case and the corresponding results match.
However, the AMD prescription \cite{Ashtekar:1999jx} for computing the mass of
a hairy spacetime is not suitable when the scalar field breaks the asymptotic
anti-de Sitter invariance \cite{Anabalon:2014fla}.

\subsection{Exact hairy solutions and triple-trace deformations}

\label{exact3}

As a concrete example, we discuss the boundary conditions and some holographic
properties of the exact solutions of \cite{Acena:2013jya, Anabalon:2013eaa}.
We consider the following scalar potential, which for some particular values
of the parameter $\Upsilon$ it becomes the one of a truncation of $\omega
$-deformed gauged $\mathcal{N}=8$ supergravity \cite{Anabalon:2013eaa,
Guarino:2013gsa, Tarrio:2013qga}:
\begin{align}
V(\phi)  &  =\frac{\Lambda(\nu^{2}-4)}{6\kappa\nu^{2}}\biggl{[}\frac{\nu
-1}{\nu+2}e^{-\phi l_{\nu}(\nu+1)}+\frac{\nu+1}{\nu-2}e^{\phi l_{\nu}(\nu
-1)}+4\frac{\nu^{2}-1}{\nu^{2}-4}e^{-\phi l_{\nu}}\biggr{]}\\
&  +\frac{\Upsilon}{\kappa\nu^{2}}\biggl{[}\frac{\nu-1}{\nu+2}\sinh{\phi
l_{\nu}(\nu+1)}-\frac{\nu+1}{\nu-2}\sinh{\phi l_{\nu}(\nu-1)}+4\frac{\nu
^{2}-1}{\nu^{2}-4}\sinh{\phi l_{\nu}}\biggr{]}\nonumber
\end{align}

Using the metric ansatz (\ref{Ansatz}), the equations of motion can be
integrated for the conformal factor \cite{Anabalon:2013sra, Anabalon:2013qua,
Acena:2012mr, Acena:2013jya}:
\begin{equation}
\Omega(x)=\frac{\nu^{2}x^{\nu-1}}{\eta^{2}(x^{\nu}-1)^{2}}%
\end{equation}
where $\Upsilon$, $\nu$, $\kappa$ and $\Lambda=-3 l^{-2}$ are parameters of
the potential and $\eta$ is an integration constant. All of them characterize
the hairy solution. With this choice of the conformal factor, it is
straightforward to obtain the expressions for the scalar field
\begin{equation}
\phi(x)=l_{\nu}^{-1}\ln{x}%
\end{equation}
and metric function
\begin{equation}
f(x)=\frac{1}{l^{2}}+\Upsilon\biggl{[}\frac{1}{\nu^{2}-4}-\frac{x^{2}}{\nu
^{2}}\biggl{(}1+\frac{x^{-\nu}}{\nu-2}-\frac{x^{\nu}}{\nu+2}%
\biggr{)}\biggr{]}+\frac{x}{\Omega(x)}%
\end{equation}
where $l_{\nu}^{-1}=\sqrt{(\nu^{2}-1)/2\kappa}$.

We would like to point out that this potential is symmetric under
$\nu\rightarrow-\nu$. For $x=1$, which corresponds to the boundary, we can
show that the theory has a standard AdS vacuum $2\kappa V(\phi=0)=2\Lambda$.
In the limit $\nu=1$, one gets $l_{\nu}\rightarrow\infty$ and $\phi
\rightarrow0$ so that the Schwarzschild-AdS black hole is smoothly obtained.

To compare with the results presented in the previous section, we should work
with the canonical coordinates of AdS. Let us discuss the branch
$x\in(1,\infty)$ for which the scalar field is positively defined. We change
the $r$-coordinate so that the function in front of the transversal section,
$d\Sigma_{k}$, has the following fall-off:
\begin{equation}
\Omega(x)=r^{2}+O(r^{-3})
\end{equation}
This choice is motivated by the fact that the term $O(r^{-2})$ generates a
lineal term in the fall-off of $\Omega$. The first three subleading terms are
\begin{equation}
x=1+\frac{1}{\eta r}+\frac{m}{r^{3}}+\frac{n}{r^{4}}+\frac{p}{r^{5}}+O(r^{-6})
\end{equation}
and they can be computed by considering the expansion around $r=\infty$:
\begin{equation}
\Omega(x)=r^{2}-\frac{24m\eta^{3}+\nu^{2}-1}{12\eta^{2}}-\frac{24n\eta^{4}%
-\nu^{2}+1}{12\eta^{3} r}+\frac{720m^{2}\eta^{6}-480p\eta^{5}+\nu^{4}%
-20\nu^{2}+19}{240\eta^{4} r^{2}}+O(r^{-3})
\end{equation}
After a straightforward computation we obtain
\begin{equation}
x=1+\frac{1}{\eta r}-\frac{(\nu^{2}-1)}{24\eta^{3}r^{3}}\biggl{[}1-\frac
{1}{\eta r}- \frac{9(\nu^{2}-9)}{80\eta^{2}r^{2}}\biggr{]}+O(r^{-6})
\end{equation}
and the following asymptotic expansions for the metric functions:
\begin{equation}
-g_{tt}=f(x)\Omega(x)=\frac{r^{2}}{l^{2}}+1+\frac{\Upsilon+3\eta^{2} }%
{3\eta^{3} r}+O(r^{-3})
\end{equation}
\begin{equation}
g_{rr}=\frac{\Omega(x)\eta^{2}}{f(x)}\biggl{(}\frac{dx}{dr}\biggr{)}=\frac
{l^{2}}{r^{2}}- \frac{l^{4}}{r^{4}}-\frac{l^{2}(\nu^{2}-1)}{4\eta^{2} r^{4}}-
\frac{l^{2}(3\eta^{2} l^{2}+\Upsilon l^{2}-\nu^{2}+1)}{3\eta^{3} r^{5}%
}+O(r^{-6})
\end{equation}
The asymptotic expansion of the scalar field becomes in these coordinates
\begin{equation}
\phi(x)= l_{\nu}^{-1}\ln{x}= \frac{1}{l_{\nu}\eta r}-\frac{1}{2l_{\nu}\eta
^{2}r^{2}}-\frac{\nu^{2}-9}{24\eta^{3}r^{3}}+O(r^{-4})
\end{equation}
and then, in the standard notation, we obtain $\alpha=1/l_{\nu}\eta$,
$\beta=-1/2l_{\nu}\eta^{2}$. Both modes are normalizable and, since
$\beta=C\alpha^{2}$ with $C=-l_{\nu}/2$, the conformal symmetry in the
boundary is preserved. Now, we can easily compute the Hamiltonian mass of the
system as was proposed in \cite{Anabalon:2014fla}
\begin{equation}
M=\sigma\biggl{[}\frac{\mu}{\kappa}+\frac{1}{l^{2}}\biggl{(}W-\frac{\alpha}{3}\frac
{dW}{d\alpha}\biggr{)}\biggr{]}
\end{equation}
and by considering $W=-l_{\nu}\alpha^{3}/6$, $\sigma=4\pi$, and $l_{\nu}%
^{-1}=\sqrt{(\nu^{2}-1)/2\kappa}$ we obtain
\begin{equation}
M=-\frac{\sigma}{\kappa}\biggl{(}\frac{3\eta^{2}+\Upsilon}{3\eta^{3}}\biggr{)}
\end{equation}
that matches the holographic mass.

Let us end up this subsection with the interpretation of these hairy solutions
within AdS/CFT duality. That is, since $W=-l_{\nu}\alpha^{3}/6$, they
correspond to adding a triple trace deformation to the boundary action as in
(\ref{triple}) (similar examples can be found in \cite{Hertog:2004dr,
Hertog:2004rz}):
\begin{equation}
I_{CFT}\rightarrow I_{CFT} +\frac{l_{\nu}}{6} \int d^{3}x\mathcal{O}^{3}%
\end{equation}
For different hairy black holes, which are characterized by the hairy
parameter $\nu$, the relation between $\alpha$ and $\beta$ does not change and
so there are triple trace deformations, but with different couplings $l_{\nu
}/6$.

\section{Conclusions}

Since the paper is self-contained and each section contains detailed
computations and interpretations, we would like only to present some general
conclusions and possible future directions.

The counterterm method \cite{Balasubramanian:1999re}, which was obtained in
the context of AdS/CFT duality, is by now a textbook example of regularizing
the gravitational action. Initially it was proposed for asymptotically AdS
solutions \cite{deHaro:2000xn,Bianchi:2001kw,Skenderis:2002wp} and then it was
generalized to asymptotically flat solutions
\cite{Astefanesei:2005ad,Mann:2005yr,Mann:2006bd,Astefanesei:2006zd,
Marolf:2006bk,Astefanesei:2009mc,Astefanesei:2009wi,Astefanesei:2010bm} and
even dS solutions
\cite{Balasubramanian:2001nb,Ghezelbash:2001vs,Ghezelbash:2002vz,
Astefanesei:2003gw}, though in the last two cases it is fair to say that there
is no valid holographic interpretation generally accepted. Interestingly, this
method provides the quasilocal stress tensor and conserved charges in a very
similar way with the well understood holography of asymptotically AdS spacetimes.

When the theory contains scalar fields, there is a diversity of mixed boundary
conditions that can be imposed, in particular boundary conditions that break
the conformal symmetry of the boundary. The `holographic renormalization'
method \cite{Skenderis:2000in,deHaro:2000xn,Bianchi:2001kw,Skenderis:2002wp}
that uses the Fefferman-Graham expansion was generalized for the mixed
boundary conditions that correspond to the non-logarithmic branch of solutions
in \cite{Papadimitriou:2007sj}.

In this work we have constructed explicit covariant counterterms that are
similar with the ones proposed by Balasubramanian and Kraus
\cite{Balasubramanian:1999re} and generalized this method for theories (moduli
potentials) that contain also the logarithmic branch of solutions. To
construct these counterterms we were guided by the Hamiltonian method that
provides the correct boundary conditions (in particular, the fall-off of the
scalar field) so that the conserved charges are finite. We did also check that
the variational principle for the gravitational action is well-posed. It may
then not be surprising that the holographic mass matches the Hamiltonian mass
for all the boundary conditions. However, when comparing with AMD formalism
there is a drastic change when the boundary conditions do not preseve the
conformal symmetry and, as was shown in \cite{Anabalon:2014fla}, the AMD mass
is not suitable for this case.

As future directions we would like to consider counterterms for other
conformal masses of the scalar field and for theories in higher dimensions. It
will be useful, if possible, to provide a general algorithm for constructing
the counterterms by using the Hamiltonian method --- it is not at all clear if
that is possible for gravity solutions that are asymptotically dS. For
the extremal black holes, there exists a different method to compute the
conserved charges, the entropy function formalism proposed by Sen
\cite{Sen:2005wa,Sen:2005iz,Sen:2007qy} (for spinning black holes it was
generalized in \cite{Astefanesei:2006dd} and in the context of AdS/CFT
duality, see e.g. \cite{Astefanesei:2007vh,
Morales:2006gm,Astefanesei:2008wz,Astefanesei:2010dk,Astefanesei:2011pz}).
However, this method provides the charges by using the near horizon geometry
data and, when there is a non-trivial RG flow in theories with scalars turned
on, it will be interesting to compare the conserved charges computed at the
horizon with the ones obtained at the boundary by the counterterm method. The
counterterm method was already used in \cite{Anabalon:2015ija} to study the
phase diagram of a general class of hairy black holes with spherical horizon
geometry and we also plan to study the phase transitions for the case $k=0$
\cite{Dumitru} when a `hairy' AdS soliton can be constructed.

A different perspective, which naturally arises when scalar fields and gravity
interact, is the classical issue of hairy black holes. Indeed, it was very
early shown that in asymptotically flat spacetimes and when the scalar field
potential is convex, the only spherically symmetric black hole is the
Schwarzschild solution \cite{Bekenstein:1972ny, Bekenstein:1972ky}, which was
later generalized to non-negative self interactions \cite{Heusler:1992ss,
Sudarsky:1995zg}, for a recent review see \cite{Herdeiro:2015waa}. These
no-hair theorems were not expected to hold when, asymptotically, there is a
non-trivial cosmological constant. The numerical existence of asymptotically
AdS black holes has been verified in a number of papers \cite{Hertog:2004dr,
Torii:2001pg, Sudarsky:2002mk} (a number exact hairy black holes has been
found when the scalar field mass is $m^{2}=-2l^{-2}$ \cite{Martinez:2004nb,
Kolyvaris:2009pc, Anabalon:2012sn, Anabalon:2012ta, Gonzalez:2013aca,
Acena:2013jya, Anabalon:2013eaa, Feng:2013tza}). Some of these black holes are
linearly stable \cite{Torii:2001pg,Anabalon:2015vda}. Another interesting
direction is on boson star solutions and the relation with the instabilities
of some AdS solutions (and AdS itself)
\cite{Astefanesei:2003qy,Astefanesei:2003rw,Buchel:2015rwa,Bizon:2011gg,Buchel:2013uba}%
.

We hope to report in the near future some progress in these directions.

\section{Acknowledgments}

DA would like to thank Stefan Theisen for interesting discussions on conformal
anomalies. DA and DC acknowledge the hospitality of the Albert Einstein
Institute in Potsdam during the last stages of this research. This work has
been done with partial support from the Fondecyt grants 11121187, 1120446,
1121031,1130658, and 1141073 and by the Newton-Picarte Grants DPI20140053 and
DPI20140115. D. C. thanks CONICYT for a Ph.D. scholarship. The Centro de
Estudios Cient\'{\i}ficos (CECs) is funded by the Chilean Government through
the Centers of Excellence Base Financing Program of Conicyt.


\begin{thebibliography}{99}                                                                                               %


\bibitem {Maldacena:1997re}J.~M.~Maldacena, ``The Large N limit of
superconformal field theories and supergravity,''
Adv.\ Theor.\ Math.\ Phys.\ \textbf{2}, 231 (1998) [hep-th/9711200].


\bibitem {Henningson:1998gx}M.~Henningson and K.~Skenderis, ``The Holographic
Weyl anomaly,'' JHEP \textbf{9807}, 023 (1998) [hep-th/9806087].

\bibitem {Balasubramanian:1999re}V.~Balasubramanian and P.~Kraus,
\textquotedblleft A Stress tensor for Anti-de Sitter
gravity,\textquotedblright\ Commun.\ Math.\ Phys.\ \textbf{208} (1999) 413
[hep-th/9902121].


\bibitem {Ishibashi:2004wx}A.~Ishibashi and R.~M.~Wald, \textquotedblleft
Dynamics in nonglobally hyperbolic static space-times. 3. Anti-de Sitter
space-time,\textquotedblright\ Class.\ Quant.\ Grav.\ \textbf{21} (2004) 2981
[hep-th/0402184].


\bibitem {BF}P.~Breitenlohner and D.~Z.~Freedman, \textquotedblleft Positive
Energy in anti-De Sitter Backgrounds and Gauged Extended
Supergravity,\textquotedblright\ Phys.\ Lett.\ B \textbf{115} (1982)
197;  P.~Breitenlohner and D.~Z.~Freedman, \textquotedblleft Stability
in Gauged Extended Supergravity,\textquotedblright\ Annals Phys.\ \textbf{144}
(1982) 249.


\bibitem {Henneaux:2002wm}M.~Henneaux, C.~Mart\'{\i}nez, R.~Troncoso and
J.~Zanelli, ``Black holes and asymptotics of 2+1 gravity coupled to a scalar
field,'' Phys.\ Rev.\ D \textbf{65}, 104007 (2002) [hep-th/0201170].


\bibitem {Barnich:2002pi}G.~Barnich, \textquotedblleft Conserved charges in
gravitational theories: Contribution from scalar fields,\textquotedblright%
\ gr-qc/0211031.


\bibitem {Henneaux:2004zi}M.~Henneaux, C.~Mart\'{\i}nez, R.~Troncoso and
J.~Zanelli, \textquotedblleft Asymptotically anti-de Sitter spacetimes and
scalar fields with a logarithmic branch,\textquotedblright\ Phys.\ Rev.\ D
\textbf{70}, 044034 (2004) [hep-th/0404236].


\bibitem {Hertog:2004dr}T.~Hertog and K.~Maeda, \textquotedblleft Black holes
with scalar hair and asymptotics in N = 8 supergravity,\textquotedblright%
\ JHEP \textbf{0407}, 051 (2004) [hep-th/0404261].


\bibitem {Hertog:2004ns}T.~Hertog and G.~T.~Horowitz, \textquotedblleft
Designer gravity and field theory effective potentials,\textquotedblright%
\ Phys.\ Rev.\ Lett.\ \textbf{94} (2005) 221301 [hep-th/0412169].


\bibitem {Henneaux:2006hk}M.~Henneaux, C.~Mart\'{\i}nez, R.~Troncoso and
J.~Zanelli, \textquotedblleft Asymptotic behavior and Hamiltonian analysis of
anti-de Sitter gravity coupled to scalar fields,\textquotedblright\ Annals
Phys.\ \textbf{322} (2007) 824 [hep-th/0603185].


\bibitem {Amsel:2006uf}A.~J.~Amsel and D.~Marolf, \textquotedblleft Energy
Bounds in Designer Gravity,\textquotedblright\ Phys.\ Rev.\ D \textbf{74},
064006 (2006) [Erratum-ibid.\ D \textbf{75}, 029901 (2007)] [hep-th/0605101].


\bibitem {Anabalon:2014fla}A.~Anabal\'on, D.~Astefanesei and C.~Mart\'{\i}nez,
``Mass of asymptotically anti-de Sitter hairy spacetimes,'' Phys.\ Rev.\ D
\textbf{91}, no. 4, 041501 (2015) [arXiv:1407.3296 [hep-th]].


\bibitem {Witten:2001ua}E.~Witten, \textquotedblleft Multitrace operators,
boundary conditions, and AdS / CFT correspondence,\textquotedblright%
\ hep-th/0112258.


\bibitem {Aharony:2015afa}O.~Aharony, G.~Gur-Ari and N.~Klinghoffer, ``The
Holographic Dictionary for Beta Functions of Multi-trace Coupling Constants,''
JHEP \textbf{1505}, 031 (2015) [arXiv:1501.06664 [hep-th]].


\bibitem {Acena:2012mr}A.~Acena, A.~Anabal\'on and D.~Astefanesei, ``Exact
hairy black brane solutions in $AdS_{5}$ and holographic RG flows,''
Phys.\ Rev.\ D \textbf{87}, no. 12, 124033 (2013) [arXiv:1211.6126 [hep-th]].


\bibitem {Acena:2013jya}A.~Ace\~{n}a, A.~Anabal\'{o}n, D.~Astefanesei and
R.~Mann, ``Hairy planar black holes in higher dimensions,'' JHEP
\textbf{1401}, 153 (2014) [arXiv:1311.6065 [hep-th]].

\bibitem {Anabalon:2013sra}A.~Anabal\'on and D.~Astefanesei, ``On attractor
mechanism of AdS$_{4}$ black holes,'' Phys.\ Lett.\ B \textbf{727}, 568 (2013)
[arXiv:1309.5863 [hep-th]].


\bibitem{Fan:2015ykb} 
  Z.~Y.~Fan and B.~Chen,
  ``Exact formation of hairy planar black holes,''
  arXiv:1512.09145 [hep-th].

\bibitem {Skenderis:2000in}K.~Skenderis, ``Asymptotically Anti-de Sitter
space-times and their stress energy tensor,'' Int.\ J.\ Mod.\ Phys.\ A
\textbf{16}, 740 (2001) [hep-th/0010138].


\bibitem {Skenderis:2002wp}K.~Skenderis, ``Lecture notes on holographic
renormalization,'' Class.\ Quant.\ Grav.\ \textbf{19}, 5849 (2002)
[hep-th/0209067].


\bibitem {Brown:1992br}J.~D.~Brown and J.~W.~York, Jr., ``Quasilocal energy
and conserved charges derived from the gravitational action,'' Phys.\ Rev.\ D
\textbf{47} (1993) 1407 [gr-qc/9209012].


\bibitem {Papadimitriou:2007sj}I.~Papadimitriou, \textquotedblleft Multi-Trace
Deformations in AdS/CFT: Exploring the Vacuum Structure of the Deformed
CFT,\textquotedblright\ JHEP \textbf{0705} (2007) 075 [hep-th/0703152].

\bibitem{Aparicio:2012yq} 
  J.~Aparicio, D.~Grumiller, E.~Lopez, I.~Papadimitriou and S.~Stricker,
  JHEP {\bf 1305}, 128 (2013)
  doi:10.1007/JHEP05(2013)128
  [arXiv:1212.3609 [hep-th]].

\bibitem{Nojiri:1998dh}
  S.~Nojiri and S.~D.~Odintsov,
  ``Conformal anomaly for dilaton coupled theories from AdS / CFT correspondence,''
  Phys.\ Lett.\ B {\bf 444} (1998) 92
  [hep-th/9810008].

\bibitem{Nojiri:2000kh}
  S.~Nojiri, S.~D.~Odintsov and S.~Ogushi,
  ``Finite action in d-5 gauged supergravity and dilatonic conformal anomaly for dual quantum field theory,''
  Phys.\ Rev.\ D {\bf 62} (2000) 124002
  [hep-th/0001122].

\bibitem {Ashtekar:1999jx}A.~Ashtekar and S.~Das, ``Asymptotically Anti-de
Sitter space-times: Conserved quantities,'' Class.\ Quant.\ Grav.\ \textbf{17}%
, L17 (2000) [hep-th/9911230]; A.~Ashtekar and A.~Magnon,
``Asymptotically anti-de Sitter space-times,''
Class.\ Quant.\ Grav.\ \textbf{1} (1984) L39.

\bibitem {Chow:2013gba}D.~D.~K.~Chow and G.~Comp\`{e}re, ``Dyonic AdS black
holes in maximal gauged supergravity,'' Phys.\ Rev.\ D \textbf{89}, no. 6,
065003 (2014) [arXiv:1311.1204 [hep-th]].


\bibitem {Anabalon:2013eaa}A.~Anabal\'on and D.~Astefanesei, ``Black holes in
$\omega$-defomed gauged $N=8$ supergravity,'' Phys.\ Lett.\ B \textbf{732},
137 (2014) [arXiv:1311.7459 [hep-th]].

\bibitem {Dall'Agata:2012bb}G.~Dall'Agata, G.~Inverso and M.~Trigiante,
``Evidence for a family of SO(8) gauged supergravity theories,''
Phys.\ Rev.\ Lett.\ \textbf{109} (2012) 201301 [arXiv:1209.0760 [hep-th]].


\bibitem {Tarrio:2013qga}J.~Tarr\'io and O.~Varela, ``Electric/magnetic
duality and RG flows in AdS$_{4}$/CFT$_{3}$,'' JHEP \textbf{1401} (2014) 071
[arXiv:1311.2933 [hep-th]].


\bibitem {Dibitetto:2014sfa}G.~Dibitetto, A.~Guarino and D.~Roest, ``Lobotomy
of Flux Compactifications,'' JHEP \textbf{1405} (2014) 067 [arXiv:1402.4478
[hep-th]].


\bibitem {Gallerati:2014xra}A.~Gallerati, H.~Samtleben and M.~Trigiante, ``The
$\mathcal{N}>2 $ supersymmetric AdS vacua in maximal supergravity,'' JHEP
\textbf{1412} (2014) 174 [arXiv:1410.0711 [hep-th]].




\bibitem {Emparan:1999pm}R.~Emparan, C.~V.~Johnson and R.~C.~Myers, ``Surface
terms as counterterms in the AdS / CFT correspondence,'' Phys.\ Rev.\ D
\textbf{60}, 104001 (1999) [hep-th/9903238].


\bibitem {Balasubramanian:1998sn}V.~Balasubramanian, P.~Kraus and
A.~E.~Lawrence, ``Bulk versus boundary dynamics in anti-de Sitter
space-time,'' Phys.\ Rev.\ D \textbf{59}, 046003 (1999) [hep-th/9805171].
Balasubramanian:1998sn, Balasubramanian:1998de

\bibitem {Balasubramanian:1998de}V.~Balasubramanian, P.~Kraus, A.~E.~Lawrence
and S.~P.~Trivedi, ``Holographic probes of anti-de Sitter space-times,''
Phys.\ Rev.\ D \textbf{59}, 104021 (1999) [hep-th/9808017].

\bibitem {Gubser:1998bc}S.~S.~Gubser, I.~R.~Klebanov and A.~M.~Polyakov,
``Gauge theory correlators from noncritical string theory,'' Phys.\ Lett.\ B
\textbf{428}, 105 (1998) [hep-th/9802109].


\bibitem {Witten:1998qj}E.~Witten, ``Anti-de Sitter space and holography,''
Adv.\ Theor.\ Math.\ Phys.\ \textbf{2}, 253 (1998) [hep-th/9802150].


\bibitem {Lu:2013ura}H.~L\"{u}, Y.~Pang and C.~N.~Pope, ``AdS Dyonic Black
Hole and its Thermodynamics,'' JHEP \textbf{1311}, 033 (2013) [arXiv:1307.6243
[hep-th]].

\bibitem{Lu:2014maa}
  H.~Lu, C.~N.~Pope and Q.~Wen,
  ``Thermodynamics of AdS Black Holes in Einstein-Scalar Gravity,''
  JHEP {\bf 1503} (2015) 165
  doi:10.1007/JHEP03(2015)165
  [arXiv:1408.1514 [hep-th]].

\bibitem {Gubser:1997yh}S.~S.~Gubser, I.~R.~Klebanov and A.~A.~Tseytlin,
``String theory and classical absorption by three-branes,'' Nucl.\ Phys.\ B
\textbf{499}, 217 (1997) [hep-th/9703040].

\bibitem {Gubser:1997se}S.~S.~Gubser and I.~R.~Klebanov, ``Absorption by
branes and Schwinger terms in the world volume theory,'' Phys.\ Lett.\ B
\textbf{413}, 41 (1997) [hep-th/9708005].

\bibitem {Myers:1999psa}R.~C.~Myers, ``Stress tensors and Casimir energies in
the AdS / CFT correspondence,'' Phys.\ Rev.\ D \textbf{60} (1999) 046002
[hep-th/9903203].

\bibitem {Hollands:2005wt}S.~Hollands, A.~Ishibashi and D.~Marolf,
``Comparison between various notions of conserved charges in asymptotically
AdS-spacetimes,'' Class.\ Quant.\ Grav.\ \textbf{22}, 2881 (2005)
[hep-th/0503045].


\bibitem {Regge:1974zd}T.~Regge and C.~Teitelboim, ``Role of Surface Integrals
in the Hamiltonian Formulation of General Relativity,'' Annals
Phys.\ \textbf{88} (1974) 286.


\bibitem {Banados:2005hm}M.~Ba\~nados and S.~Theisen, ``Scale invariant hairy
black holes,'' Phys.\ Rev.\ D \textbf{72}, 064019 (2005) [hep-th/0506025].


\bibitem {Gegenberg:2003jr}J.~Gegenberg, C.~Mart\'{\i}nez and R.~Troncoso, ``A
Finite action for three-dimensional gravity with a minimally coupled scalar
field,'' Phys.\ Rev.\ D \textbf{67}, 084007 (2003) [hep-th/0301190].


\bibitem {Martinez:2004nb}C.~Mart\'{\i}nez, R.~Troncoso and J.~Zanelli,
``Exact black hole solution with a minimally coupled scalar field,''
Phys.\ Rev.\ D \textbf{70} (2004) 084035 [hep-th/0406111].


\bibitem {Anabalon:2012ta}A.~Anabal\'on, ``Exact Black Holes and Universality
in the Backreaction of non-linear Sigma Models with a potential in (A)dS4,''
JHEP \textbf{1206} (2012) 127 [arXiv:1204.2720 [hep-th]].

\bibitem {Guarino:2013gsa}A.~Guarino, ``On new maximal supergravity and its
BPS domain-walls,'' JHEP \textbf{1402}, 026 (2014) [arXiv:1311.0785
[hep-th]].

\bibitem {Anabalon:2013qua}A.~Anabal\'on, D.~Astefanesei and R.~Mann, ``Exact
asymptotically flat charged hairy black holes with a dilaton potential,''
arXiv:1308.1693 [hep-th].

\bibitem {Hertog:2004rz}T.~Hertog and G.~T.~Horowitz, ``Towards a big crunch
dual,'' JHEP \textbf{0407}, 073 (2004) [hep-th/0406134].

\bibitem {deHaro:2000xn}S.~de Haro, S.~N.~Solodukhin and K.~Skenderis,
``Holographic reconstruction of space-time and renormalization in the AdS /
CFT correspondence,'' Commun.\ Math.\ Phys.\ \textbf{217}, 595 (2001)
[hep-th/0002230].


\bibitem {Bianchi:2001kw}M.~Bianchi, D.~Z.~Freedman and K.~Skenderis,
``Holographic renormalization,'' Nucl.\ Phys.\ B \textbf{631}, 159 (2002)
[hep-th/0112119].


\bibitem {Mann:2005yr}R.~B.~Mann and D.~Marolf, ``Holographic renormalization
of asymptotically flat spacetimes,'' Class.\ Quant.\ Grav.\ \textbf{23}, 2927
(2006) [hep-th/0511096].

\bibitem {Astefanesei:2005ad}D.~Astefanesei and E.~Radu, ``Quasilocal
formalism and black ring thermodynamics,'' Phys.\ Rev.\ D \textbf{73}, 044014
(2006) [hep-th/0509144].


\bibitem {Mann:2006bd}R.~B.~Mann, D.~Marolf and A.~Virmani, ``Covariant
Counterterms and Conserved Charges in Asymptotically Flat Spacetimes,''
Class.\ Quant.\ Grav.\ \textbf{23}, 6357 (2006) [gr-qc/0607041].


\bibitem {Astefanesei:2006zd}D.~Astefanesei, R.~B.~Mann and C.~Stelea, ``Note
on counterterms in asymptotically flat spacetimes,'' Phys.\ Rev.\ D
\textbf{75}, 024007 (2007) [hep-th/0608037].


\bibitem {Marolf:2006bk}D.~Marolf, ``Asymptotic flatness, little string
theory, and holography,'' JHEP \textbf{0703}, 122 (2007) [hep-th/0612012].


\bibitem {Astefanesei:2009mc}D.~Astefanesei, M.~J.~Rodriguez and S.~Theisen,
``Quasilocal equilibrium condition for black ring,'' JHEP \textbf{0912}, 040
(2009) [arXiv:0909.0008 [hep-th]].


\bibitem {Astefanesei:2009wi}D.~Astefanesei, R.~B.~Mann, M.~J.~Rodriguez and
C.~Stelea, ``Quasilocal formalism and thermodynamics of asymptotically flat
black objects,'' Class.\ Quant.\ Grav.\ \textbf{27}, 165004 (2010)
[arXiv:0909.3852 [hep-th]].


\bibitem {Astefanesei:2010bm}D.~Astefanesei, M.~J.~Rodriguez and S.~Theisen,
``Thermodynamic instability of doubly spinning black objects,'' JHEP
\textbf{1008}, 046 (2010) [arXiv:1003.2421 [hep-th]].


\bibitem {Balasubramanian:2001nb}V.~Balasubramanian, J.~de Boer and D.~Minic,
``Mass, entropy and holography in asymptotically de Sitter spaces,''
Phys.\ Rev.\ D \textbf{65}, 123508 (2002) [hep-th/0110108].


\bibitem {Ghezelbash:2001vs}A.~M.~Ghezelbash and R.~B.~Mann, ``Action, mass
and entropy of Schwarzschild-de Sitter black holes and the de Sitter / CFT
correspondence,'' JHEP \textbf{0201}, 005 (2002) [hep-th/0111217].


\bibitem {Ghezelbash:2002vz}A.~M.~Ghezelbash, D.~Ida, R.~B.~Mann and
T.~Shiromizu, ``Slicing and brane dependence of the (A)dS / CFT
correspondence,'' Phys.\ Lett.\ B \textbf{535}, 315 (2002) [hep-th/0201004].


\bibitem {Astefanesei:2003gw}D.~Astefanesei, R.~B.~Mann and E.~Radu,
``Reissner-Nordstrom-de Sitter black hole, planar coordinates and dS / CFT,''
JHEP \textbf{0401}, 029 (2004) [hep-th/0310273].

\bibitem {Sen:2005iz}A.~Sen, ``Entropy function for heterotic black holes,''
JHEP \textbf{0603}, 008 (2006) [hep-th/0508042].

\bibitem {Sen:2005wa}A.~Sen, ``Black hole entropy function and the attractor
mechanism in higher derivative gravity,'' JHEP \textbf{0509}, 038 (2005)
[hep-th/0506177].

\bibitem {Sen:2007qy}A.~Sen, ``Black Hole Entropy Function, Attractors and
Precision Counting of Microstates,'' Gen.\ Rel.\ Grav.\ \textbf{40}, 2249
(2008) [arXiv:0708.1270 [hep-th]].

\bibitem {Astefanesei:2006dd}D.~Astefanesei, K.~Goldstein, R.~P.~Jena, A.~Sen
and S.~P.~Trivedi, ``Rotating attractors,'' JHEP \textbf{0610}, 058 (2006)
[hep-th/0606244].

\bibitem {Astefanesei:2008wz}D.~Astefanesei, N.~Banerjee and S.~Dutta,
``(Un)attractor black holes in higher derivative AdS gravity,'' JHEP
\textbf{0811}, 070 (2008) [arXiv:0806.1334 [hep-th]].


\bibitem {Astefanesei:2007vh}D.~Astefanesei, H.~Nastase, H.~Yavartanoo and
S.~Yun, ``Moduli flow and non-supersymmetric AdS attractors,'' JHEP
\textbf{0804}, 074 (2008) [arXiv:0711.0036 [hep-th]].


\bibitem {Morales:2006gm}J.~F.~Morales and H.~Samtleben, ``Entropy function
and attractors for AdS black holes,'' JHEP \textbf{0610}, 074 (2006)
[hep-th/0608044].


\bibitem {Astefanesei:2010dk}D.~Astefanesei, N.~Banerjee and S.~Dutta,
``Moduli and electromagnetic black brane holography,'' JHEP \textbf{1102}, 021
(2011) [arXiv:1008.3852 [hep-th]].


\bibitem {Astefanesei:2011pz}D.~Astefanesei, N.~Banerjee and S.~Dutta, ``Near
horizon data and physical charges of extremal AdS black holes,''
Nucl.\ Phys.\ B \textbf{853}, 63 (2011) [arXiv:1104.4121 [hep-th]].


\bibitem {Anabalon:2015ija}A.~Anabal\'on, D.~Astefanesei and D.~Choque, ``On
the thermodynamics of hairy black holes,'' Phys.\ Lett.\ B \textbf{743}, 154
(2015) [arXiv:1501.04252 [hep-th]].


\bibitem {Dumitru}A. Anabal\'on, D. Astefanesei, and D. Choque, (in preparation)

\bibitem {Bekenstein:1972ny}J.~D.~Bekenstein, ``Transcendence of the law of
baryon-number conservation in black hole physics,''
Phys.\ Rev.\ Lett.\ \textbf{28} (1972) 452.

\bibitem {Bekenstein:1972ky}J.~D.~Bekenstein, ``Nonexistence of baryon number
for black holes. ii,'' Phys.\ Rev.\ D \textbf{5} (1972) 2403.

\bibitem {Heusler:1992ss}M.~Heusler, ``A No hair theorem for selfgravitating
nonlinear sigma models,'' J.\ Math.\ Phys.\ \textbf{33} (1992) 3497.

\bibitem {Sudarsky:1995zg}D.~Sudarsky, ``A Simple proof of a no hair theorem
in Einstein Higgs theory,,'' Class.\ Quant.\ Grav.\ \textbf{12} (1995) 579.

\bibitem {Herdeiro:2015waa}C.~A.~R.~Herdeiro and E.~Radu, ``Asymptotically
flat black holes with scalar hair: a review,'' arXiv:1504.08209 [gr-qc].


\bibitem {Torii:2001pg}T.~Torii, K.~Maeda and M.~Narita, ``Scalar hair on the
black hole in asymptotically anti-de Sitter space-time,'' Phys.\ Rev.\ D
\textbf{64} (2001) 044007.


\bibitem {Sudarsky:2002mk}D.~Sudarsky and J.~A.~Gonzalez, ``On black hole
scalar hair in asymptotically anti-de Sitter space-times,'' Phys.\ Rev.\ D
\textbf{67} (2003) 024038 [gr-qc/0207069].

\bibitem {Kolyvaris:2009pc}T.~Kolyvaris, G.~Koutsoumbas, E.~Papantonopoulos
and G.~Siopsis, ``A New Class of Exact Hairy Black Hole Solutions,''
Gen.\ Rel.\ Grav.\ \textbf{43} (2011) 163 [arXiv:0911.1711 [hep-th]].


\bibitem {Anabalon:2012sn}A.~Anabal\'on, F.~Canfora, A.~Giacomini and
J.~Oliva, ``Black Holes with Primary Hair in gauged N=8 Supergravity,'' JHEP
\textbf{1206} (2012) 010 [arXiv:1203.6627 [hep-th]].


\bibitem {Gonzalez:2013aca}P.~A.~Gonz\'alez, E.~Papantonopoulos, J.~Saavedra
and Y. V\'asquez, ``Four-Dimensional Asymptotically AdS Black Holes with
Scalar Hair,'' JHEP \textbf{1312} (2013) 021 [arXiv:1309.2161 [gr-qc]].


\bibitem {Feng:2013tza}X.~H.~Feng, H.~Lu and Q.~Wen, ``Scalar Hairy Black
Holes in General Dimensions,'' Phys.\ Rev.\ D \textbf{89} (2014) 4, 044014
[arXiv:1312.5374 [hep-th]].


\bibitem {Anabalon:2015vda}A.~Anabal\'on, D.~Astefanesei and J.~Oliva, ``Hairy
Black Hole Stability in AdS, Quantum Mechanics on the Half-Line and
Holography,'' JHEP \textbf{1510}, 068 (2015) [arXiv:1507.05520 [hep-th]].

\bibitem {Astefanesei:2003qy}D.~Astefanesei and E.~Radu, ``Boson stars with
negative cosmological constant,'' Nucl.\ Phys.\ B \textbf{665}, 594 (2003)
[gr-qc/0309131].


\bibitem {Bizon:2011gg}P.~Bizon and A.~Rostworowski, ``On weakly turbulent
instability of anti-de Sitter space,'' Phys.\ Rev.\ Lett.\ \textbf{107},
031102 (2011) [arXiv:1104.3702 [gr-qc]].


\bibitem {Buchel:2013uba}A.~Buchel, S.~L.~Liebling and L.~Lehner, ``Boson
stars in AdS spacetime,'' Phys.\ Rev.\ D \textbf{87}, no. 12, 123006 (2013)
[arXiv:1304.4166 [gr-qc]].


\bibitem {Astefanesei:2003rw}D.~Astefanesei and E.~Radu, ``Rotating boson
stars in (2+1)-dimenmsions,'' Phys.\ Lett.\ B \textbf{587}, 7 (2004)
[gr-qc/0310135].


\bibitem {Buchel:2015rwa}A.~Buchel, ``AdS boson stars in string theory,''
arXiv:1510.08415 [hep-th].
\end{thebibliography}
\end{document}